\documentclass[prb,aps,amsmath,amssymb,showpacs,twocolumn]{revtex4}

\usepackage{bm,color,float,dcolumn,amsmath,amssymb,amsfonts,amsthm,graphicx,hyperref,subfigure,dsfont}

\newcommand{\be}{\begin{equation}}
\newcommand{\ee}{\end{equation}}

\newcommand{\ba}{\begin{eqnarray}}
\newcommand{\ea}{\end{eqnarray}}

\begin{document}

\title{Creating and manipulating non-Abelian anyons in cold atom systems using auxiliary bosons}
\author{Yuhe Zhang,$^1$ G. J. Sreejith,$^2$ and J. K. Jain$^1$}

\affiliation{
   $^{1}$Department of Physics, 104 Davey Lab, Pennsylvania State University, University Park, PA 16802, USA}
   \affiliation{$^2$Max Planck Institute for Physics of Complex Systems, N\"{o}thnitzer St 38, 01187  Dresden, Germany }

\date{\today}

\begin{abstract} 

The possibility of realizing bosonic fractional quantum Hall effect in ultra-cold atomic systems suggests a new route to producing and manipulating anyons, by introducing auxiliary bosons of a different species that capture quasiholes and thus inherit their non-trivial braiding properties. States with localized quasiholes at any desired locations can be obtained by annihilating the auxiliary bosons at those locations. We explore how this method can be used to generate non-Abelian quasiholes of the Moore-Read Pfaffian state for bosons at filling factor $\nu=1$. We show that a Hamiltonian with an appropriate three-body interaction can produce two-quasihole states in two distinct fusion channels of the topological ``qubit." Characteristics of these states that are related to the non-Abelian nature can be probed and verified by a measurement of the effective relative angular momentum of the auxiliary bosons, which is directly related to their pair distribution function. Moore-Read states of more than two quasiholes can also be produced in a similar fashion.  We investigate some issues related to the experimental feasibility of this approach, in particular, how large the systems should be for a realization of this physics and to what extent this physics carries over to systems with the more standard two-body contact interaction.

\end{abstract}

\maketitle

\section{Introduction}

Emergent particles in two-dimensional systems can, in principle, transcend the dichotomy of boson and fermion. The existence of the exotic statistics stems from the fact that winding one particle around another in two dimensions is topologically inequivalent to a process in which none of the particles move at all. Thus braiding particles does not necessarily bring the system back to the same state. Particles obeying fractional braid statistics\cite{Leinaas77,Wilczek82,Wilczek90} are called ``anyons." For these particles, the wave function acquires, when one particle winds around another, a phase that is a non-integral multiple of $2\pi$. Since the product of phase factors is a commutative operation, this type of statistics is called Abelian braid statistics, associated with one-dimensional representations of the braid group. Non-Abelian statistics can arise if there is a degenerate set of wave functions for quasiparticles at fixed positions. Braiding of some particles around the others then corresponds to a unitary matrix transformation in the space of such states. If such matrices are non-commutative, the particles are said to exhibit non-Abelian statistics. Systems whose excitations satisfy non-Abelian braid statistics are of interest both for a demonstration of this physics, and because of their potential application in storage and processing of quantum information~\cite{Nayak08}. If such degenerate states with quasiparticles at fixed locations are separated from the rest of the spectrum by an energy gap, then for sufficiently slow perturbations, dynamics is restricted to within such a degenerate subspace. Any local perturbation has no non-trivial matrix elements within this degenerate subspace, and therefore braiding of quasiparticles is essentially the only way to perform nontrivial unitary operations on this subspace at low energies, immunizing the system against decoherence.~\cite{Kitaev03} 

Of the many theoretical proposals for the realization of non-Abelian anyons~\cite{Moore91,Read99,Jolicoeur07,Scarola02b, Dassarma06, Gurarie05, Kitaev06,Levin05a,Levin05b,Fendley05,Freedman05}, the fractional quantum Hall (FQH) state of electrons at filling $\nu=5/2$ is perhaps the most promising. It is believed to be described by the Moore-Read (MR) Pfafian wave function~\cite{Moore91}, which represents a chiral p-wave paired state\cite{Read00} of composite fermions\cite{Jain89,Jain07,Jain15}. Numerical studies have shown support  for this interpretation of the $5/2$ FQH state: the exact Coulomb ground state at $\nu = 1/2$ in the second Landau level (LL) is in the same universality class as the MR state for systems of up to $18$ electrons,\cite{Morf98, Rezayi00, Storni10}
and the MR wave function has a significantly lower energy than the spin-polarized or the spin-unpolarized composite-fermion Fermi sea state in the second LL~\cite{Park98b}. Abrikosov vortices of a chiral p-wave superconductor carry Majorana fermions and obey non-Abelian braid statistics \cite{Read00}, and it has been proposed that the quasihole excitations of the MR state also exhibit non-Abelian statistics~\cite{Moore91}, with $2n$ quasiholes at fixed positions spanning a $2^{n-1}$-dimensional degenerate space~\cite{Nayak96}.

Quantum Hall interferometry~\cite{Chamon97, Fradkin98, DasSarma05, Stern06, Bonderson06a, Bonderson06b, Chung06} in the solid state devices has been the best explored among the proposals for demonstration of non-Abelian statistics. It relies on the idea that the transport of quasiparticles in the edge-currents of a Hall bar is equivalent to braiding them around the quasiparticles in the bulk, and the phases acquired in such braiding processes can be detected through interference between currents along different paths. Experimental progress has been made toward the detection of such interferences~\cite{Willett09, Willett10, Willett12, Willett13a, Willett13b, An11}, but the interpretation of the experiments in terms of non-Abelian anyons has not been unambiguous. Theoretical work has suggested conceptual difficulties stemming primarily from the `volatility' in the location of the edge and significant effort has gone into understanding all the effects that determine the edge fluctuations and into extracting signals of non-Abelian statistics.~\cite{Rosenow07,Rosenow09,Halperin11,Keyserlingk14,Zhang09,Ofek10} 

In this work, we show that some of these problems can be overcome, in principle, in cold atom systems. Bosonic atoms are neutral, but they behave as charged bosons in a magnetic field when subjected to rotation, because, formally, rotation plays the same role as a perpendicular magnetic field~\cite{Ho01,Cooper08,Viefers08}. The filling factor is then given by the ratio of boson density to vortex density. 
At sufficiently rapid rotation (i.e. low filling factor), neutral bosons can be driven into the strongly correlated regime of the quantum Hall effect~\cite{Wilkin98,Wilkin00,Cooper01,Regnault03,Chang05b,Cooper08,Viefers08,Wu13,Furukawa13,Regnault13,Meyer14}. Another way to mimic magnetic field in neutral atomic systems is to generate artificial gauge potentials using atom-light interaction.~\cite{Dalibard11,Goldman14}
Quantized vortices have been produced with various experimental techniques in atomic Bose \cite{Matthews99,Madison00,Haljan01,Raman01,Lin09,Lin11} and Fermi gases~\cite{Zwierlein05}, and extremely high vorticity has been achieved~\cite{Aboshaeer01,Bretin04,Schweikhard04}. Progress towards producing the fractional quantum Hall effect (FQHE) conditions has been made by employing an adiabatic pathway to transfer the atom clusters from zero angular momentum into a ground state of any higher total angular momentum $L$.~\cite{Popp04,Gemelke10} In addition to the schemes described above, there have been many other proposals~\cite{Sorensen05,Hafezi07,Kapit10, Nielsen13, Nielsen14,Tang11, Sun11, Neupert11,Yao13,Cooper13} aiming to realize FQHE or analogous behavior in lattice models, with ultracold atoms confined in optical lattice~\cite{Jaksch05,Lewenstein07,Bloch08}. In this paper, we focus on continuum atomic gases in two dimensions. While FQHE has not been produced in such systems so far, we will assume that it is possible. We will also assume that it is possible to engineer arbitrary 2- or 3-body interactions in cold atom systems. A method for realization of 3-body interaction has been proposed by Roncaglia {\em et al.}\cite{Roncaglia10} recently with implementation requirements within present or planned technologies.

One advantage of cold atoms systems is that it is possible to introduce auxiliary bosons of a different species and manipulate them.\cite{Stamper-Kurn98,Stenger98,Burke98,Bloch01,Barrett01} We have shown in a previous paper~\cite{Zhang14} that for appropriate interactions, the auxiliary bosons bind quasiholes of the quantum Hall droplet, and thus serve as ``place holders'' for quasiholes. They also inherit the fractional statistics of the quasiholes. One can thus imagine manipulating the quasiholes by manipulating the auxiliary bosons. In this article, we construct a 3-body interaction that accomplishes the same for quasiholes obeying non-Abelian braid statistics. 
We demonstrate that a specific choice of 3-body interaction can be used to prepare states of two quasiholes in either of the two different ``fusion channels" labeled ${\bf 1}$ or $\psi$, which differ in their topological character and serve as the elementary qubit of a putative topological quantum computer based on the FQHE. (Swapping the two states can be achieved by braiding a third quasihole around one of the two quasiholes.) 
The two states can be distinguished through the different Berry phases produced when one quasihole encircles another.

Another advantage of cold atom systems is that signatures of non-Abelian statistics can be seen in the effective relative angular momentum of a pair of auxiliary bosons~\cite{Zhang14}. We show that the fractional part of the relative angular momentum, which can be deduced from the pair correlation function of the pair of auxiliary bosons, corresponds to the fractional Berry phase accumulated when one quasihole is taken around another. Using explicit numerical calculation of the pair correlation function, we show that the fractional angular momenta indeed correctly reflect the fractional Berry phases associated with different fusion channels as predicted~\cite{Nayak08} by conformal field theory (CFT). This demonstration of non-trivial statistics, while still being at the level of a thought-experiment, has the advantage that all the physics happen away from the edge of the droplet. Another aspect of non-Abelian excitations are the multiplicities of states with excitations at fixed locations. Such degeneracies are reflected in the states with auxiliary bosons carrying quasiholes.

While the situation is relatively clear, at least theoretically, for a model 3-body interaction, such an interaction will not be the easiest to implement in cold atom systems. Fortunately, it has been found~\cite{Wilkin00, Cooper01, Regnault03, Regnault07, Chang05b} that even bosons with 2-body contact interaction exhibit the chiral p-wave paired Pfaffian state at $\nu=1$. Our detailed studies show, however, that the applicability of most of the above ideas does not extend to a system of bosons with 2-body contact interaction, at least for system sizes available in our studies. This is perhaps related to the studies~\cite{Toke06a} of the 5/2 state for electrons that show that the excitations of the 5/2 state for the Coulomb interaction do not have a one-to-one correspondence to the excitations of the 3-body interaction for small systems amenable to numerical studies. We note that {\em quasiparticles} in the MR state have also been described theoretically in a way analogous to the quasiholes.~\cite{Hansson09a, Hansson09b,Sreejith11b} This work focuses on the quasiholes of the MR state.

There have been many other theoretical approaches~\cite{Paredes01,Kapit12,JuliaDiaz12,Grass14,Cooper15} towards the goal of engineering anyons in quantum Hall regime and probing their peculiar statistics with the help of ultra-cold atoms and optical control. Paredes {\em et al.}~\cite{Paredes01} proposed to create quasiholes by focusing laser beams onto the atomic cloud and braiding them by adiabatically moving the laser beam. Juli\'a-D\'{i}az {\em et al.}\cite{JuliaDiaz12} and Gra\ss{} {\em et al.}~\cite{Grass14} demonstrated by calculating the Berry phase that Abelian fractional statistics manifest in very small systems, consistent with our previous results~\cite{Zhang14}. Kapit {\em et al.}~\cite{Kapit12} reported on a numerical experiment to test the non-Abelian braiding properties of lattice bosons, which could be implemented with cold atoms in a deep optical lattice and an artificial gauge field. Different from the above approaches that require time-dependent potential to braid the quasiparticles, Cooper and Simon \cite{Cooper15} proposed to demonstrate Haldane's fractional exclusion statistics of quasiholes in the bosonic Laughlin state through  spectroscopic measurements. 

The organization of the paper is as follows. In Sec. \ref{sec: background}, we review the relevant ideas regarding the MR Pfaffian state and its quasihole excitations; and construct wave functions describing the 2-quasihole qubit. We then construct a 3-body Hamiltonian in a system of $N$ majority bosons and two auxiliary bosons in Sec. \ref{sec: Hamiltonian} which produce the 2-quasihole qubit states as its exact eigenstates. Assuming a system in such a qubit state, we demonstrate in Sec. \ref{sec: statistics} how the non-Abelian statistics can be seen by measuring the pair correlation function of the pair of auxiliary bosons. We further study how to produce $2n$-quasihole states in Sec. \ref{sec: manyquasihole} and discuss the experimental feasibility of our proposals in cold atom systems in Sec. \ref{sec: exp}. Finally we conclude in Sec. \ref{sec: conclusion}.

\section{Background \label{sec: background}}

We begin with a review of the MR Pfaffian ground state wave function~\cite{Moore91}, the many-quasihole states, as well as the explicit qubit representation for 4-quasihole states~\cite{Nayak96}. We show that these wave functions can be expressed in a bipartite form. In particular, two distinct 2-quasihole wave functions of the bipartite form are shown to be related to the 4-quasihole qubit states with two of the quasiholes sent to infinity.

\subsection{Moore-Read state and quasiholes}

Moore and Read~\cite{Moore91} constructed a set of trial wave functions for various FQH states using the analogy between the correlators in various CFTs and the incompressible FQH states. The MR state of the form
\begin{equation}
\begin{aligned}
\Psi _{1/m}^{\text{Pf}} =  \prod _{i<j} (z_{i} - z_{j} )^{m} \mbox{Pf} \left(\frac{1}{z_{i}-z_{j}}\right) e^{ -\sum_{i} \frac{|z_{i}|^{2}}{4\ell^{2}}}
\label{Pfaffian}
\end{aligned}
\end{equation}
has a filling fraction $\nu=\frac{1}{m}$ where $m$ is even for fermions and odd for bosons. Here $z_{i}=x_{i}+\imath y_{i}$ label the complex coordinates of the particles, and $\ell=\sqrt{\hbar c/eB}$ is the magnetic length. This state represents a chiral p-wave superconducting state of composite fermions\cite{Read00}. It is also the highest density exact zero-energy eigenstate of a 3-body repulsive contact interaction, discussed in the following.

Moore and Read also suggested half-flux quantum excitations of the Pfaffian state described by the trial wave function:
\begin{equation}
\begin{gathered}
\Psi_{\text{2qh}} \propto \prod_{i<j}(z_{i}-z_{j})^{m}\times \text{Pf}(M) \times e^{ -\sum_{i} \frac{|z_{i}|^{2}}{4\ell^{2}}}\\
\text{where } M_{ij} = \frac{ (z_{i}-\eta_{1})(z_{j}-\eta_{2})+(z_{i}-\eta_{2})(z_{j}-\eta_{1}) }{z_{i}-z_{j}}.
\end{gathered}
\label{2qhMR}
\end{equation}
This state has two quasiholes localized at $\eta_{1}=x_{1}^{\prime}+\imath y_{1}^{\prime}$ and $\eta_{2}=x_{2}^{\prime}+ \imath y_{2}^{\prime}$, each with a charge $\frac{1}{2m}$. (In the following we will suppress the Gaussian factor $e^{ -\sum_{i} \frac{|z_{i}|^{2}}{4\ell^{2}}}$ for ease of notation.)

The wave function for the state consisting of $2n$ quasiholes can be constructed by modifying the argument of the Pfaffian in Eq. (\ref{2qhMR}) as follows~\cite{Nayak96}:
\begin{equation}
M_{ij} = \frac{ (z_{i}-\eta_{\alpha})(z_{i}-\eta_{\beta})...(z_{j}-\eta_{\rho})(z_{j}-\eta_{\sigma})... + (i \leftrightarrow j) }{z_{i}-z_{j}}.
\label{manyqh}
\end{equation}
In this expression, the $2n$ quasiholes have been divided into two groups of $n$ each (i.e. $\alpha$, $\beta$ $\cdots$ and $\rho$, $\sigma$ $\cdots$). There are $\frac{(2n)!}{2(n!)^2}$ ways of making such a division. Nayak and Wilczek~\cite{Nayak96} demonstrated that they span a space of only $2^{n-1}$ linearly independent states, consistent with the predictions from CFT. This degeneracy can be understood by noting that each quasihole supports a majorana fermion (a half composite fermion). Each pair of Majorana fermions in such a state can fuse in two distinct ways associated with opposite fermion parity, thus describing a Hilbert space of dimension $2^{n}$. Taking into account the fixed parity of the overall wave function, this produces $2^{n-1}$ distinct states. 

A set of linearly independent basis states $\{\varphi_{F,\mathbf{\lambda}}\}$ for the space of $2n$-quasihole states has been written by Read and Rezayi~\cite{Read1996} as follows:
\begin{eqnarray}
\varphi_{F,\mathbf{\lambda}} (z,\eta)=J^m \mathcal{A}\left[\prod_{k=1}^{F}z_k ^{\lambda_k}\times  \prod_{l=1}^{\frac{N-F}{2}}\frac{\Phi(z_{F+2l-1},z_{F+2l},\eta)}{z_{F+2l-1}-z_{F+2l}}\right]\\
\text{where }\Phi (z_1,z_2,\eta) = \sum_{\tau\in S_{2n}} \prod_{l=1}^n (z_1-\eta_{\tau(2l)})(z_2-\eta_{\tau(2l-1)})\nonumber
\label{RRformOfQHs}
\end{eqnarray}
$J$ is the Jastrow factor, $\mathcal{A}$ is the antisymmetrization operator and $S_{2n}$ is the group of permutations of $\{ 1,2,3,\dots,2n \}$. For given locations $\left\{\eta_i\right\}_{i=1\to 2n}$ of the quasiholes, the possible wave functions are indexed by an integer $F$ and a sequence $\lambda$. $F$ is an integer of same parity as the number of particles $N$, such that $0\leq F\leq n$. $\lambda$ is an ordered sequence of $F$ integers such that $\lambda_i<\lambda_{i+1}$ and $0\leq\lambda_i<n$. For example, for $2n=4$ quasiholes in even $N$ system, indices $(F,\lambda)$ can take two values $(F=0,\lambda=[\;])$ and $(F=2,\lambda=[0,1])$. 
In general, for each choice of $F$, there are $\binom{n}{F}=\frac{n!}{F!(n-F)!}$ possible choices of $\lambda$.  For $2n$ quasiholes in even $N$ system, there are $\sum_{F=0,\rm{even}}^n \binom{n}{F} = 2^{n-1}$ states for a given $\{\eta_i\}$. With fixed $\{\eta_i\}$, the MR quasihole states shown in Eq. (\ref{manyqh}) span an identical space as that of $\{\varphi_{F,\mathbf{\lambda}}\}$. Each MR quasihole state can be expressed as a linear combination of the basis $\{\varphi_{F,\mathbf{\lambda}}\}$, in which the coefficients are generally functions of $\{\eta_{i}\}$. These two bases are no longer equivalent when $\{\eta_{i}\}$ represent positions of auxiliary bosons instead of localized quasiholes, as we will consider later, in which case only $\{\varphi_{F,\mathbf{\lambda}}\}$ are valid wave functions because they are symmetric in the $\eta_{i}$'s. (The MR 2-quasihole state is an exception because it is already symmetric under exchange of $\eta_{1}$ and $\eta_{2}$.)

The above discussion gives the counting for quasihole states in a system without any edge degrees of freedom or equivalently in a spherical geometry. In the presence of an `open' edge, the parity constraint is relaxed because the edge can accommodate a majorana fermion. This is in addition to the purely bosonic excitations which are fully localized at the edge. In the above mentioned basis $\varphi_{F,\lambda}$, the upper limits on values of $F$ and $\lambda$ arise from a hard angular momentum cut off representing the edge. These constraints are again relaxed in the presence of edge degrees of freedom, thus resulting in increased degeneracy.

Multiple degenerate states with fixed quasihole positions are necessary for non-Abelian statistics. The simplest case with such a degeneracy has four quasiholes, whose states form a two-dimensional Hilbert space.\cite{Nayak96} A basis set $\{\Psi^1,\Psi^\psi\}$ can be chosen following Nayak and Wilczek which has simple braiding properties:
\begin{equation}
\begin{aligned}
\Psi^{(1,\psi)} = &\prod_{\alpha<\beta}^{4} \eta_{\alpha \beta}^{\frac{1}{4m}-\frac{1}{8}} \frac{(\eta_{13}\eta_{24})^{\frac{1}{4}} }{ (1\pm\sqrt{x})^{1/2}} \\
&\times (\Psi_{(13)(24)} \pm \sqrt{x}\Psi_{(14)(23)}) e^{-  \sum_{\alpha=1}^{4} \frac{|\eta_{\alpha}|^{2}}{8m\ell^{2}}} .
\label{Nayak}
\end{aligned}
\end{equation}
where $x= \eta_{14}\eta_{23}/\eta_{13}\eta_{24}$ and $\eta_{\alpha \beta}=\eta_{\alpha}-\eta_{\beta}$. In this expression,
\begin{equation}
\begin{aligned}
\Psi_{(\alpha \beta)(\rho \sigma)} =& \prod_{i<j}(z_{i}-z_{j})^{m} \times \text{Pf} ( M )\\
\text{where } M_{ij}=&\frac{ (z_{i}-\eta_{\alpha})(z_{i}-\eta_{\beta})(z_{j}-\eta_{\rho})(z_{j}-\eta_{\sigma}) + (i \leftrightarrow j) }{z_{i}-z_{j}}.
\label{4qh-1}
\end{aligned}
\end{equation}
An adiabatic transport of the quasihole $\eta_1$ around $\eta_2$, or of $\eta_3$ around $\eta_4$ results in an overall Abelian phase depending on the state ($1$ or $\psi$). In contrast, braiding of either $\eta_1$ or $\eta_2$ around either $\eta_3$ or $\eta_4$ results in an interchange of the states $\Psi^1$ and $\Psi^\psi$ as a result of the branch cuts in the $\sqrt{x}$ factor. If the two states represent a qubit, then such braiding operations serve as simple gates.

The states $\Psi^1$ and $\Psi^\psi$ correspond to conformal blocks of a correlator in an Ising CFT as described later in this paper. The quasihole insertion operators correspond to the $\sigma$ fields and the two states above correspond to the fusion of $\sigma(\eta_1)$ and $\sigma(\eta_2)$ into $1$ or $\psi$ channels.

\begin{figure}
\resizebox{0.41\textwidth}{!}{\includegraphics{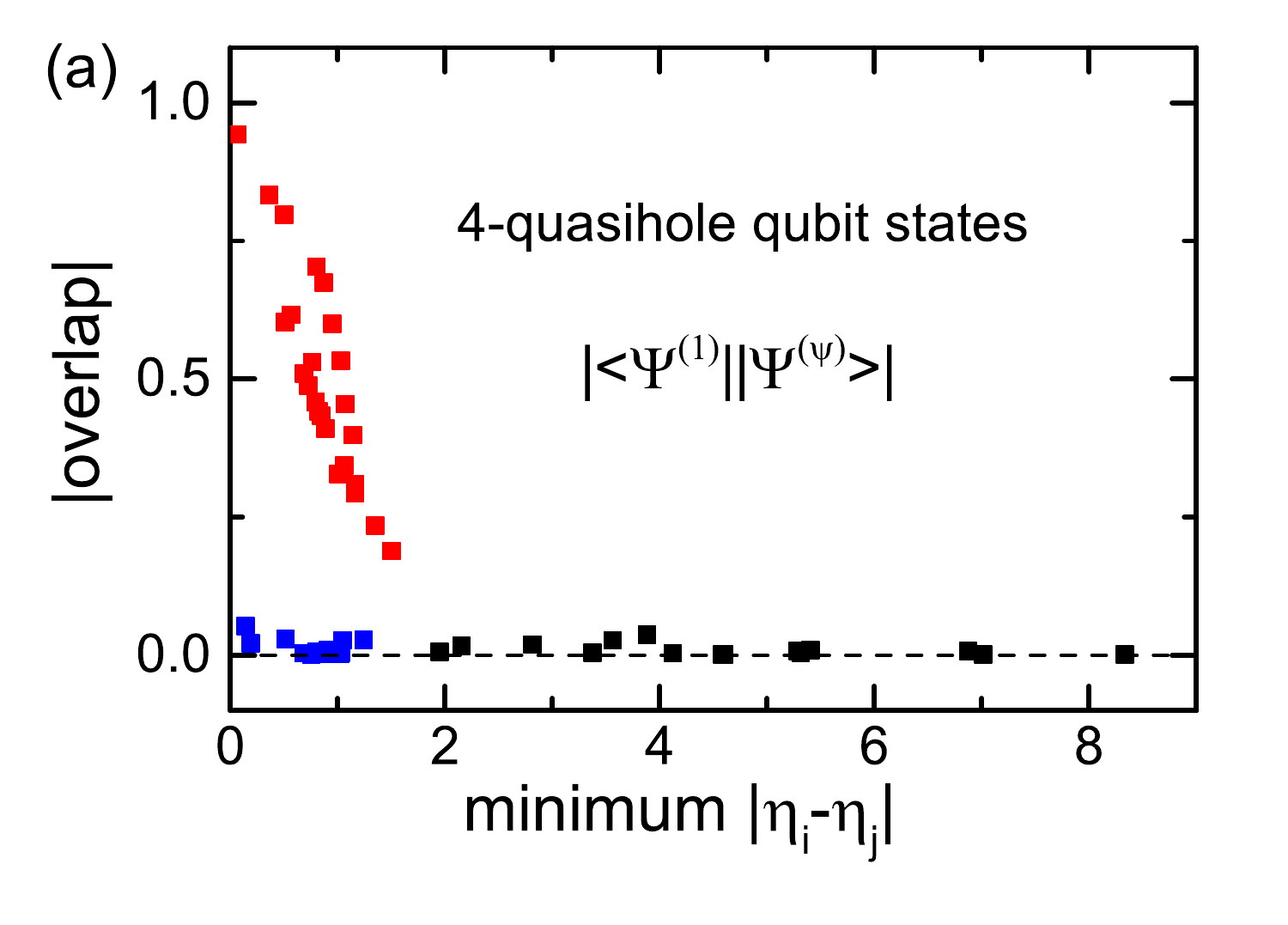}}\\
\vspace{-3.5mm}
\hspace{-0.5mm}
\resizebox{0.41\textwidth}{!}{\includegraphics{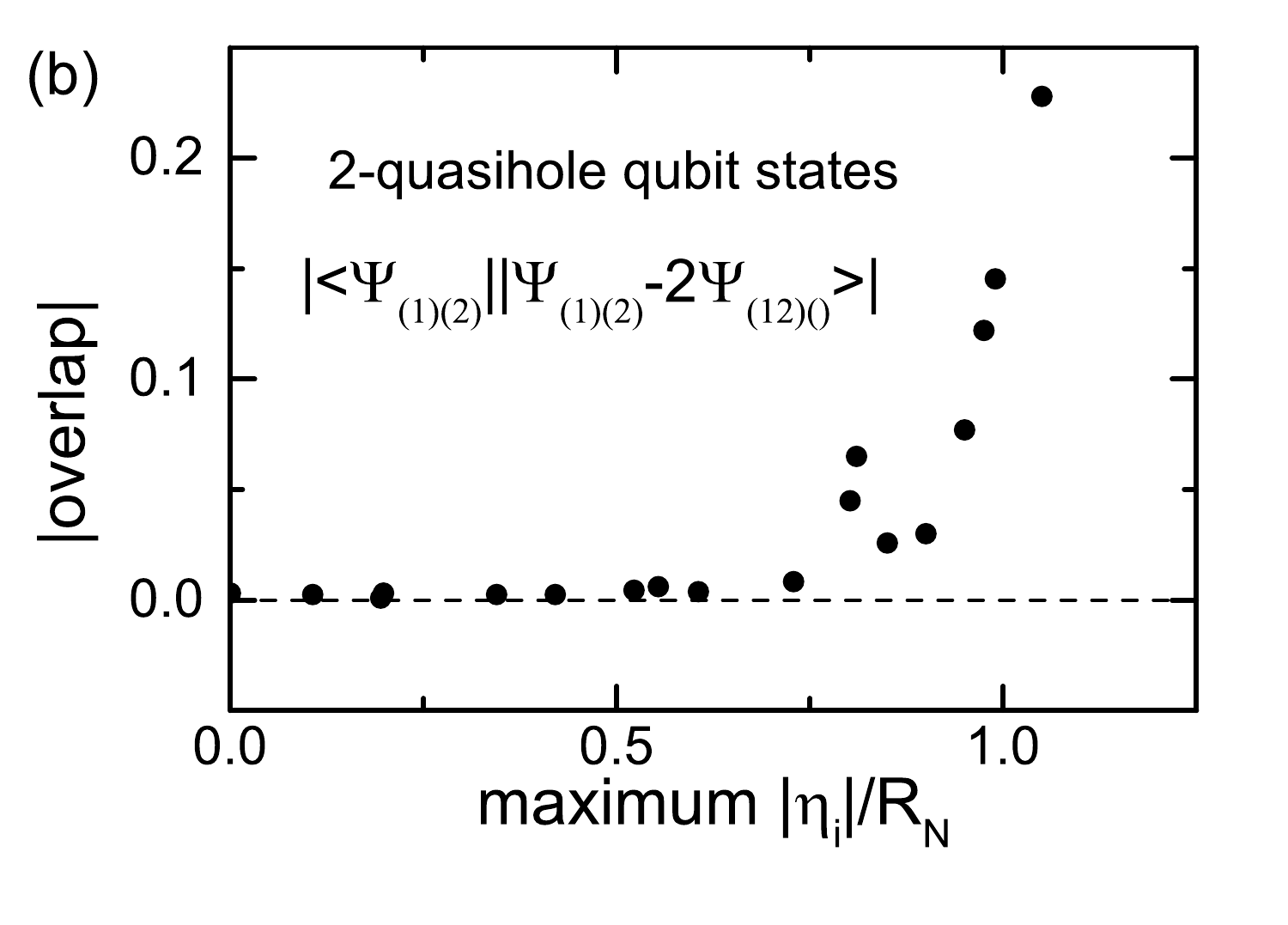}}
\vspace{-5mm}
\caption{(Color online) (a) The modulus of the overlap between the 4-quasihole qubit states $\Psi^{(1)}$ and $\Psi^{(\psi)}$ as a function of the minimum distance between two quasiholes in each quasihole-position configuration. The blue squares represent cases when the minimum distance is either $|\eta_{1} - \eta_{2}|$ or $|\eta_{3} - \eta_{4}|$, and red squares are other configurations.
(b) The modulus of the overlap between the 2-quasihole qubit states $\Psi_{(1)(2)}$ and $\Psi_{(1)(2)}-2\Psi_{(12)()}$ as a function of the maximum quasihole position $|\eta_{i}|/R_{N}$ where $R_{N}=\sqrt{2N/\nu}$ is the radius of the quantum Hall droplet. In both calculations the quasihole positions are chosen randomly, and the system size is $N=60$.}
\label{4qhoverlap}
\end{figure}

Based on a plasma analogy, Bonderson, Gurarie and Nayak demonstrated that $\Psi^{(1)}$ and $\Psi^{(\psi)}$ are orthogonal in the limit where the quasiholes are far apart from each other.~\cite{Bonderson11} Here we directly compute the overlap between the states $\Psi^{(1)}$ and $\Psi^{(\psi)}$ by Monte-Carlo technique for many different quasihole positions (chosen randomly) for a relatively large system $N=60$. Note that all the overlaps calculated in this paper are normalized, denoted as $\langle \psi_{i}||\psi_{j} \rangle =  \langle \psi_{i}|\psi_{j} \rangle / \sqrt{ \langle \psi_{i}|\psi_{i} \rangle  \langle \psi_{j}|\psi_{j} \rangle}$. Figure \ref{4qhoverlap} (a) shows the amplitude of the overlap $\langle \Psi^{(1)} ||\Psi^{(\psi)}\rangle$ as a function of the minimum distance between two quasiholes in each set of quasihole positions. The calculation shows that the overlap is zero (within numerical uncertainty) when the quasiholes are far enough from each other (e.g. $|\eta_{i} - \eta_{j}| \gtrsim 2$). When two of the quasiholes approach each other, there are two branches shown in the figure, where the red squares represent the configurations in which either $\eta_{1}$ or $\eta_{2}$ is close to either $\eta_{3}$ or $\eta_{4}$, and the blue ones indicate $\eta_{1}$ is close to $\eta_{2}$ or $\eta_{3}$ is close to $\eta_{4}$. We thus conclude that the overlap between $\Psi^{(1)}$ and $\Psi^{(\psi)}$ is essentially zero except when $\eta_{1}$ or $\eta_{2}$ approaches $\eta_{3}$ or $\eta_{4}$.

\subsection{Bipartite representation and 2-quasihole qubit \label{sec: backgroundB}}

The Pfaffian wave function can be rewritten in a bipartite form\cite{Greiter92,Cappelli99}:
\begin{equation}
\begin{gathered}
\Psi_{\nu=1/m}^{\text{Pf}} = J^{m-1} \mathcal{S} \prod_{i<j}^{N/2}(z_{2i}-z_{2j})^2  \prod_{i<j}^{N/2}(z_{2i-1}-z_{2j-1})^2,\\
J = \prod_{i<j}^N (z_i-z_j)
\end{gathered}
\label{Pfbipartite}
\end{equation}
where $\mathcal{S}$ is the symmetrizer over all the complex coordinates $\{z_i\}_{i=1\to N}$. The relation to the Pfaffian wave function can be derived with the help of an identity due to Cauchy\cite{MacDonald79}.

Our study focuses on the bosonic Pfaffian states at filling $\nu=1$ (namely $m=1$). This wave function is obtained by symmetrizing two partitions of incompressible states at filling fraction $\nu=\frac{1}{2}$. Quasihole wave functions can be produced by now inserting a single quasihole into each of these partitions as follows
\begin{equation}
\begin{gathered}
\Psi_{(1)(2)} = \mathcal{S} \left [\Phi_{\frac{1}{2},{\rm 1qh}}(z_{1},z_3,z_5\dots;\eta_1) \Phi_{\frac{1}{2},{\rm 1qh}}(z_{2},z_4,z_6\dots;\eta_2) \right]\\
\text{where }\Phi_{\frac{1}{2},{\rm 1qh}}(x_1,x_2\dots;\eta) = \prod_{i=1}^{N/2}(x_{i}-\eta) \prod_{i<j}^{N/2}(x_{i}-x_{j})^2
\end{gathered}
\label{2qh11}
\end{equation}
The subscript $(\cdots)(\cdots)$ of $\Psi$ on the left hand side, indicates the quasihole coordinates in the two partitions. Specifically, $(1)(2)$ implies that $\eta_{1}$ and $\eta_{2}$ belong to different partitions, as opposed to $(12)()$ that refers (described below) to a state in which both quasiholes reside in the same partition. We demonstrate in Appendix \ref{ap1} that this wave function $\Psi_{(1)(2)}$ is exactly the same as the MR 2-quasihole wave function written in a Pfaffian form as shown in Eq. (\ref{2qhMR}). The bipartite representation immediately suggests another 2-quasihole state in which the two quasiholes are placed in the same partition:  
\begin{equation}
\begin{aligned}
\Psi_{(12)()} = \mathcal{S} &\prod_{i=1}^{N/2}(z_{2i}-\eta_{1})(z_{2i}-\eta_{2})\prod_{i<j}^{N/2}(z_{2i}-z_{2j})^2 \\
& \times \prod_{i<j}^{N/2}(z_{2i-1}-z_{2j-1})^2.
\label{2qh20}
\end{aligned}
\end{equation}
This corresponds to the Pfaffian wave function for the 2-quasihole state (see Appendix \ref{ap1} for details):
\begin{equation}
\Psi_{(12)()} = \text{Pf} \left( \frac{ (z_{i}-\eta_{1})(z_{i}-\eta_{2})+ (i \leftrightarrow j)}{z_{i}-z_{j}} \right) \prod_{i<j}(z_{i}-z_{j})^{m}.
\label{2qh20pf}
\end{equation}

The two different 2-quasihole states $\Psi_{(1)(2)} $ and $\Psi_{(12)()} $ are related to the two different 4-quasihole states in Eq. (\ref{Nayak}). More precisely, the space spanned by these two 2-quasihole states can be obtained by taking two of the four quasiholes in Eq. (\ref{Nayak}) to infinity by setting $\eta_{3}=r$,  $\eta_{4} = r e^{i\theta}$ and taking the limit $r \rightarrow \infty$. In this limit, we find (see Appendix \ref{ap2})
\begin{equation} 
\begin{aligned}
\Psi^{(1)} & \rightarrow & \Psi^{(1)}_{\rm 2qh} &= (\eta_{1} - \eta_{2})^{1/8} \Psi_{(1)(2)}\\
\Psi^{(\psi)} & \rightarrow & \Psi^{(\psi)}_{\rm 2qh}  &= (\eta_{1} - \eta_{2})^{5/8} ( \Psi_{(1)(2)}-2\Psi_{(12)()} ).
\end{aligned}
\label{2qh1andpsi}
\end{equation}
Therefore $\Psi_{(1)(2)}$ and $\Psi_{(1)(2)}-2\Psi_{(12)()}$ form a 2-quasihole representation for the qubit states. Note that these reduced states do not transform into one another under any braiding operation; braiding $\eta_{1}$ around $\eta_{2}$ simply produces an Abelian phase. Nevertheless, the fusion properties of the $\eta_{1}$ and $\eta_{2}$ quasiholes remain the same as those in the 4-quasihole qubit system: $\eta_{1}$ and $\eta_{2}$ fuse into identity ${\bf 1}$ in  $\Psi_{(1)(2)}$, while in $\Psi_{(1)(2)}-2\Psi_{(12)()}$ they fuse into $\psi$. The fusion properties are reflected in the different Berry phases associated with braiding one quasihole around another, which we discuss in detail in Sec. \ref{sec: statistics}. The 2-quasihole states thus form a two-level system, i.e. a qubit. The state can be flipped by introducing a third quasihole that braids around one of the two quasiholes. This is exactly the picture of the constricted Hall bar as a quantum bit for making fractional quantum Hall quantum computer.~\cite{DasSarma05,Chamon97, Fradkin98}

In a finite system, the two quasiholes that are taken to large distances actually reside at the edge of the sample and thus the two wave functions differ at the edge of the system. This can also be seen when the above wave functions are realized in the spherical geometry. When placed on a sphere with a flux of $2Q=m(N-1)$, the wave functions $\Psi_{(1)(2)}$ ($\Psi_{(12)()}$) has two additional quasiholes in opposite partitions (same partitions) at the south pole. 
Additional edge excitations are possible but do not influence the properties of the bulk quasiholes.

As discussed above, the 4-quasihole qubit states are orthogonal for well-separated quasihole positions. The same is true for the 2-quasihole qubit states $\Psi_{(1)(2)}$ and $\Psi_{(1)(2)}-2\Psi_{(12)()}$. For relatively large system size ($N\geq 60$), we have performed many trial computations for different sets of $\eta_{i}$ positions, and found the overlaps are essentially zero except when a quasihole is located close to the edge of the FQH droplet. The results are shown in Fig \ref{4qhoverlap} (b). Considering that $\eta_{3,4}$ actually reside at the edge, the results indicate that the overlap is non-zero when either of $\eta_{1}$ and $\eta_{2}$ is close to $\eta_{3}$ or $\eta_{4}$, and is zero otherwise, independent of the distance $|\eta_{1} - \eta_{2}|$. This is consistent with the results for the 4-quasihole qubit states shown in Fig. \ref{4qhoverlap} (a).

\section{Producing the qubit states with 3-body interaction \label{sec: Hamiltonian}}

\subsection{Quasiholes versus auxiliary bosons}

The previous section discussed the nature of localized quasihole excitations of the Pfaffian state. In describing the quasiholes, the locations are mere parameters. In our earlier work~\cite{Zhang14}, we proposed to introduce a pair of auxiliary bosons into a background FQHE state of bosonic atoms. The auxiliary bosons capture localized vortex-like quasihole excitations of the quantum Hall state and then behave as particles with fractional braid statistics. In fact, a system with two species of particles is a natural platform for producing quasiholes, where the particles of one species (namely the auxiliary bosons) serve as ``place holders'' for the quasiholes of the other species. In other words, the quasiholes of the FQH state are enslaved by the auxiliary bosons. This idea can be illustrated with wave functions. For example, the 2-quasihole state $\Psi_{(1)(2)}$ in Eq. (\ref{2qh11}) turns into a wave function describing two species of bosons by changing the quasihole-position parameters $\eta_{i}$'s into dynamical coordinates and including a Gaussian factor for them:
\begin{equation}
\begin{aligned}
\Psi_{(1)(2)} =\mathcal{S} &\prod_{i=1}^{N/2}(z_{2i}-\eta_{1}) \prod_{i<j}^{N/2}(z_{2i}-z_{2j})^2 \prod_{i=1}^{N/2}(z_{2i-1}-\eta_{2})\\
&\times \prod_{i<j}^{N/2}(z_{2i-1}-z_{2j-1})^2 e^{ -\frac{1}{4\ell^{2}} (\sum_{i=1}^{N} |z_{i}|^{2}  +\sum_{i=1}^{2} |\eta_{i}|^{2}) }.
\label{2qh11dynamic}
\end{aligned}
\end{equation}
where $S$ symmetrizes the $z$-bosons.
Here the $\eta_{i}$-bosons and the $z_{i}$-bosons share the same magnetic length as reflected in the Gaussian factor. 
Note that $\Psi_{(1)(2)}$ is already symmetric under exchange of $\eta_{1}$ and $\eta_{2}$, making it a valid wave function even when the $\eta_{i}$'s are thought of as auxiliary boson coordinates. That is also true for $\Psi_{(12)()}$, but not for MR quasihole states with $2n>2$.
In the following, we will use $\Psi_{(1)(2)}$ and $\Psi_{(12)()}$ to denote the corresponding wave function for two species of bosons, which is slightly different from the wave function for localized quasihole states in Eqs. (\ref{2qh11}) and (\ref{2qh20}), but the meaning should be clear from the context. 

We ask if there exists a Hamiltonian that produces the  qubit states $\Psi_{(1)(2)}$ and $\Psi_{(1)(2)}-2\Psi_{(12)()}$ as its exact eigenstates. Before proceeding to that question, we first study the orthogonality and normalization properties of the 2-quasihole states.

\subsection{Orthogonality of 2-quasihole qubit states}

\begin{figure}
\resizebox{0.22\textwidth}{!}{\includegraphics{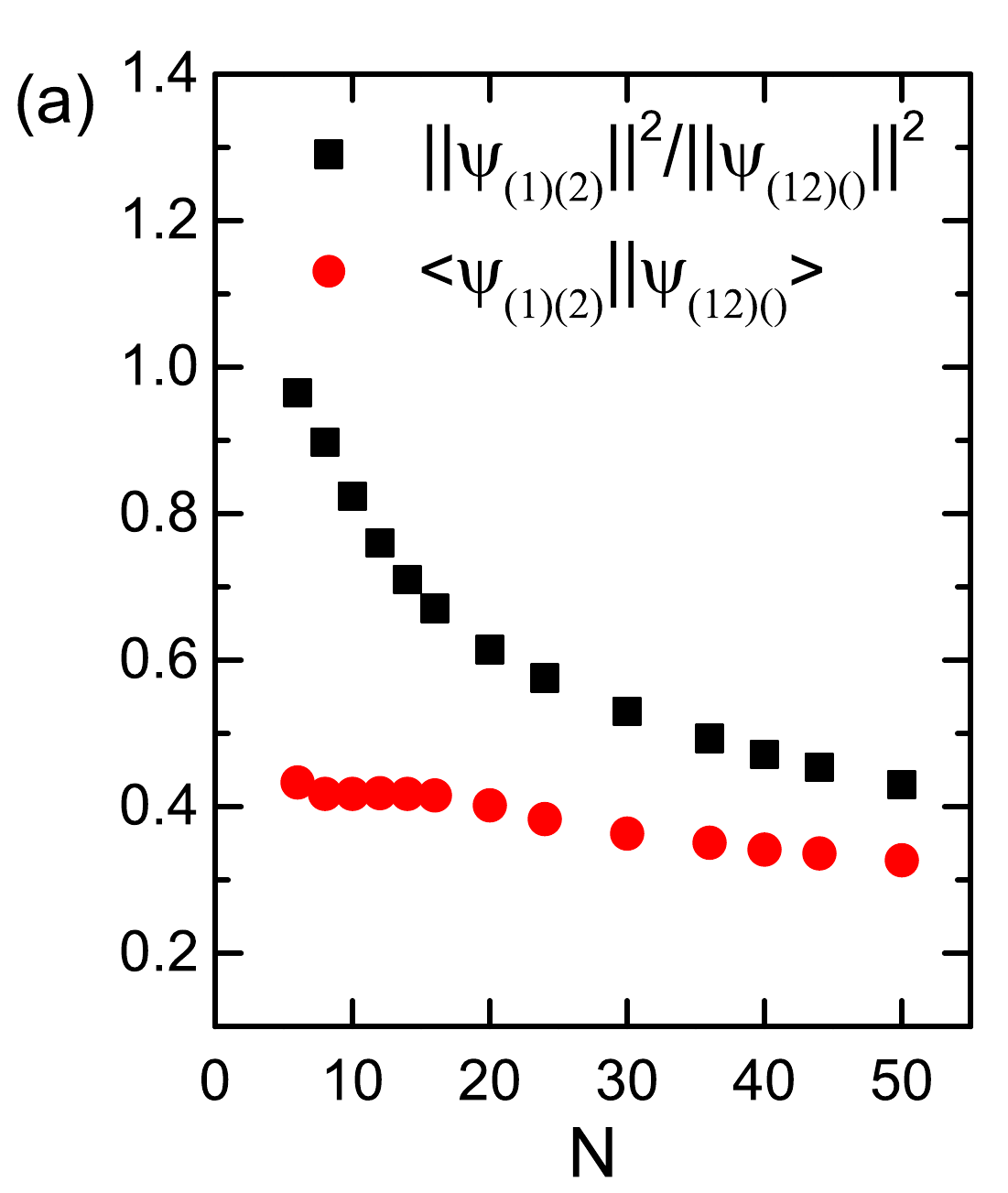}}
\hspace{-1mm}
\resizebox{0.22\textwidth}{!}{\includegraphics{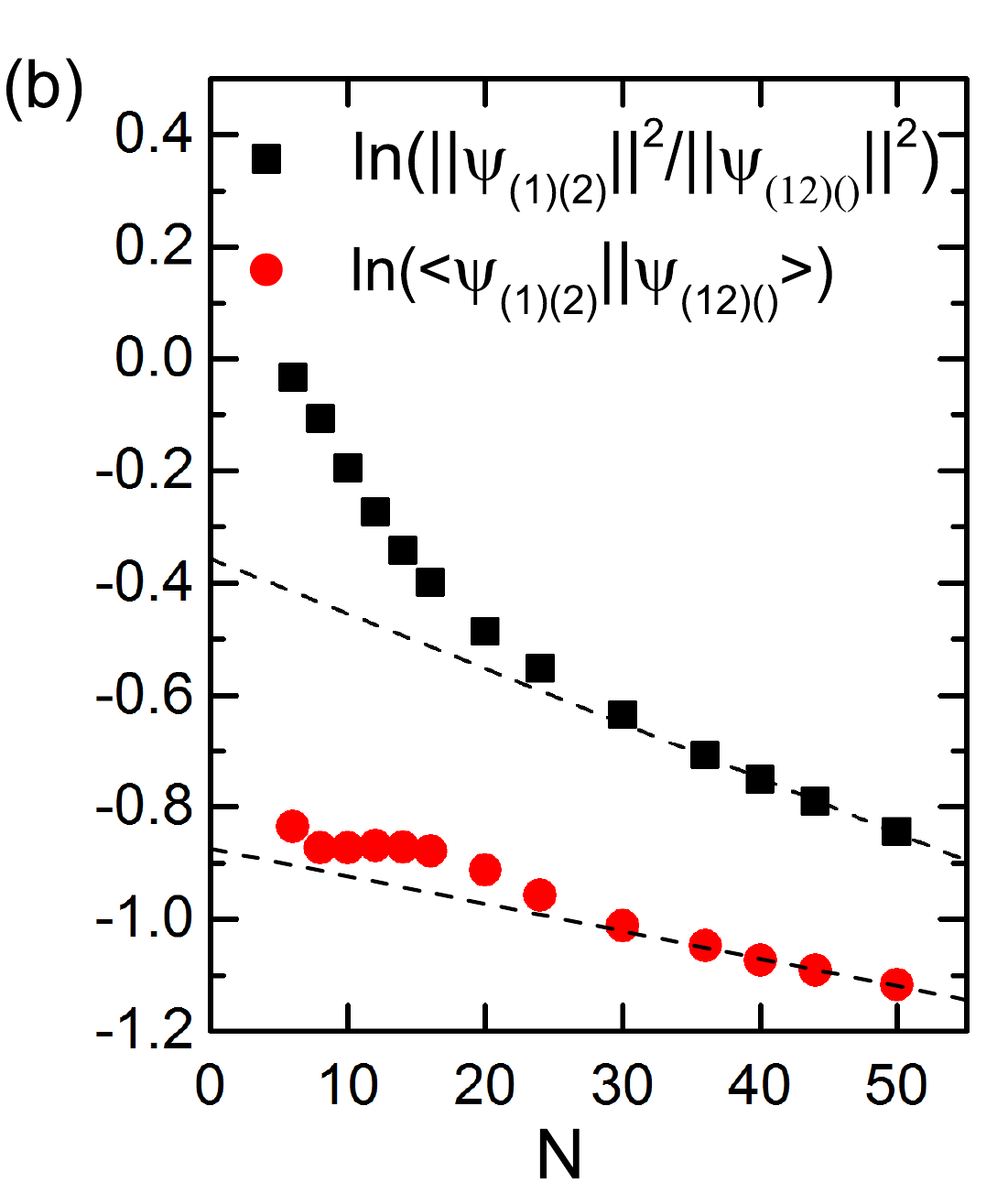}}\\
\vspace{-2mm}
\resizebox{0.37\textwidth}{!}{\includegraphics{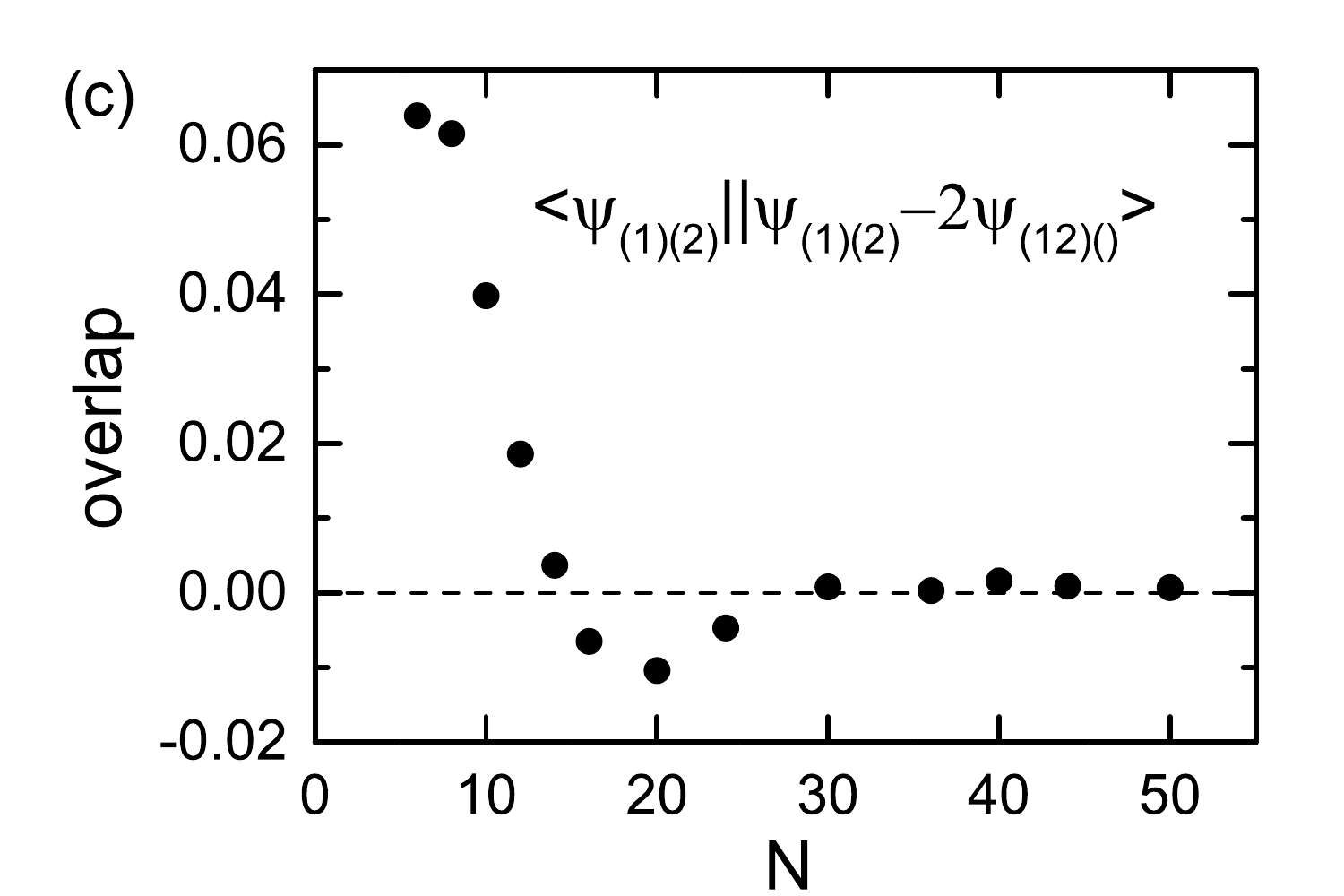}}
\caption{(Color online) (a) Black square represents the ratio of the normalization factor of the two different 2-quasihole wave functions as shown in Eq. (\ref{ratio}) and red circle represents the normalized overlap between them as shown in Eq. (\ref{overlap}) as a function of system size $N$. (b) The natural log of the corresponding quantities as a function of $N$. In panels (a) and (b), we use $||\psi_{i}||^{2} $ to denote the normalization constant $\langle \psi_{i} |\psi_{i} \rangle$. (c) The normalized overlap between the 2-quasihole qubit states $ \langle \Psi_{(1)(2)}||\Psi_{(1)(2)}-2\Psi_{(12)()} \rangle $ as a function of $N$. In all panels $\eta_{i}$'s have been taken as dynamical coordinates in Monte-Carlo calculation.}
\label{1}
\end{figure}

We have written down two different 2-quasihole states with the help of bipartite representation in the last section as $\Psi_{(1)(2)}$ and $\Psi_{(12)()}$. In order to study their properties in the thermodynamic limit, we calculate the ratio of the normalization factor of the wave functions $\Psi_{(1)(2)}$ and $\Psi_{(12)()}$
\begin{equation}
\frac{\langle \Psi_{(1)(2)} |\Psi_{(1)(2)} \rangle}{ \langle \Psi_{(12)()} |\Psi_{(12)()} \rangle} , 
\label{ratio}
\end{equation}
as well as the normalized overlap 
 \begin{equation}
\frac{\langle \Psi_{(1)(2)} |\Psi_{(12)()} \rangle}{ \sqrt{ \langle \Psi_{(1)(2)} |\Psi_{(1)(2)} \rangle   \langle \Psi_{(12)()} |\Psi_{(12)()} \rangle}}, 
\label{overlap}
\end{equation}
for different system sizes. Note that we have taken the $\eta_{i}$'s as the dynamical coordinates of the auxiliary bosons in all the calculations. Fig. \ref{1} (a) shows that the values of the two quantities decrease as the number of particles $N$ increases. The plot of the natural log of the quantity versus $N$ in Figure \ref{1}(b) indicates that both Eqs. (\ref{ratio}) and (\ref{overlap}) decrease exponentially with $N$. In the limit $N\rightarrow \infty$, $\Psi_{(12)()}$ dominates in the expression $\Psi_{(1)(2)}-2\Psi_{(12)()}$ and is orthogonal to $\Psi_{(1)(2)}$. We thus conjecture that in the thermodynamic limit, the second qubit state  $\Psi_{(1)(2)}-2\Psi_{(12)()}$ reduces to $\Psi_{(12)()}$.

We also compute the normalized overlap between the two 2-quasihole qubit states:
\begin{equation}
\frac{\langle \Psi_{(1)(2)}|\Psi_{(1)(2)}-2\Psi_{(12)()} \rangle}{   \sqrt{ \langle \Psi_{(1)(2)}|  \Psi_{(1)(2)} \rangle \langle  \Psi_{(1)(2)}-2\Psi_{(12)()} |\Psi_{(1)(2)}-2\Psi_{(12)()} \rangle }  } .
\label{overlap2qh}
\end{equation}
Figure \ref{1}(c) shows that the overlap is very small ($<0.07$) even in small system ($N=6$) and becomes essentially zero for $N \geq 30$. This result implies that $\Psi_{(1)(2)}$ and $\Psi_{(1)(2)}-2\Psi_{(12)()}$ serve as a good orthogonal basis for the 2-fold degenerate space of two quasiholes. This result will be useful in the next subsection where we show that both $\Psi_{(1)(2)}$ and $\Psi_{(1)(2)}-2\Psi_{(12)()}$ are exact zero-energy eigenstates of a certain 3-body Hamiltonian.

\subsection{Model Hamiltonians \label{sec:3bodyH}}

The Pfaffian ground state is known to be the exact zero-energy state of the repulsive 3-body interaction Hamiltonian \cite{Greiter92}:
\begin{equation}
H=V_0\sum_{i<j<k}^{N} \delta(\vec{r}_{i}-\vec{r}_{j})\delta (\vec{r}_{i}-\vec{r}_{k})
\label{3bodyH}
\end{equation}
We consider a system of two species of bosons: N $z$-bosons and two $\eta$-bosons, and take the 3-body interaction to be identical for all particles. In this sense, we write the interaction as $V=V_{zzz} + V_{zz\eta} + V_{z\eta\eta}$. We perform exact diagonalization within the basis of total angular momentum $L=N^{2}/2$, corresponding to the 2-quasihole states. Note that $\Psi_{(1)(2)}$ and $\Psi_{(12)()}$ have the same total angular momentum $L$, but different maximum single-particle angular momentum for the $z$-particle: $l_{\text{max}}^{(z)} = N-1$ and $N$ respectively. Here and in the following we consider a $N=8$ system and take the cut-off on single-particle angular momentum $l_{\text{max}}$ to be $12$ for both $z$ and $\eta$ particles. (In fact we have tested that, as long as $l_{\text{max}} \geq N$, the value of $l_{\text{max}}$ does not affect the zero-energy states obtained by diagonalizing any of the 3-body Hamiltonian we have considered here and following in the 2-quasihole case.) The exact energy spectrum shows only one zero-energy state under this interaction, which is exactly $\Psi_{(1)(2)}$.

\begin{figure}
\resizebox{0.25\textwidth}{!}{\includegraphics{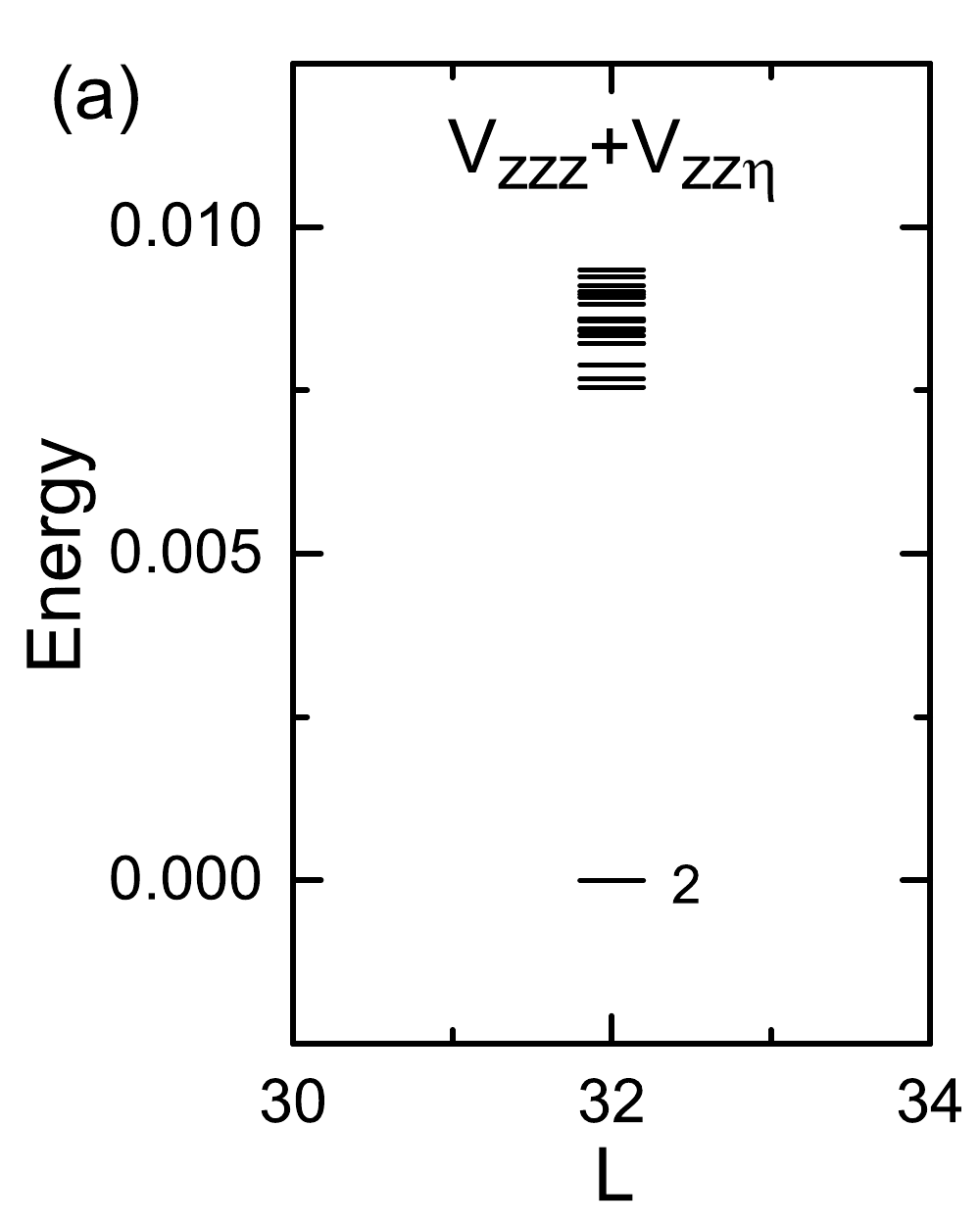}}
\hspace{-5mm}
\resizebox{0.22\textwidth}{!}{\includegraphics{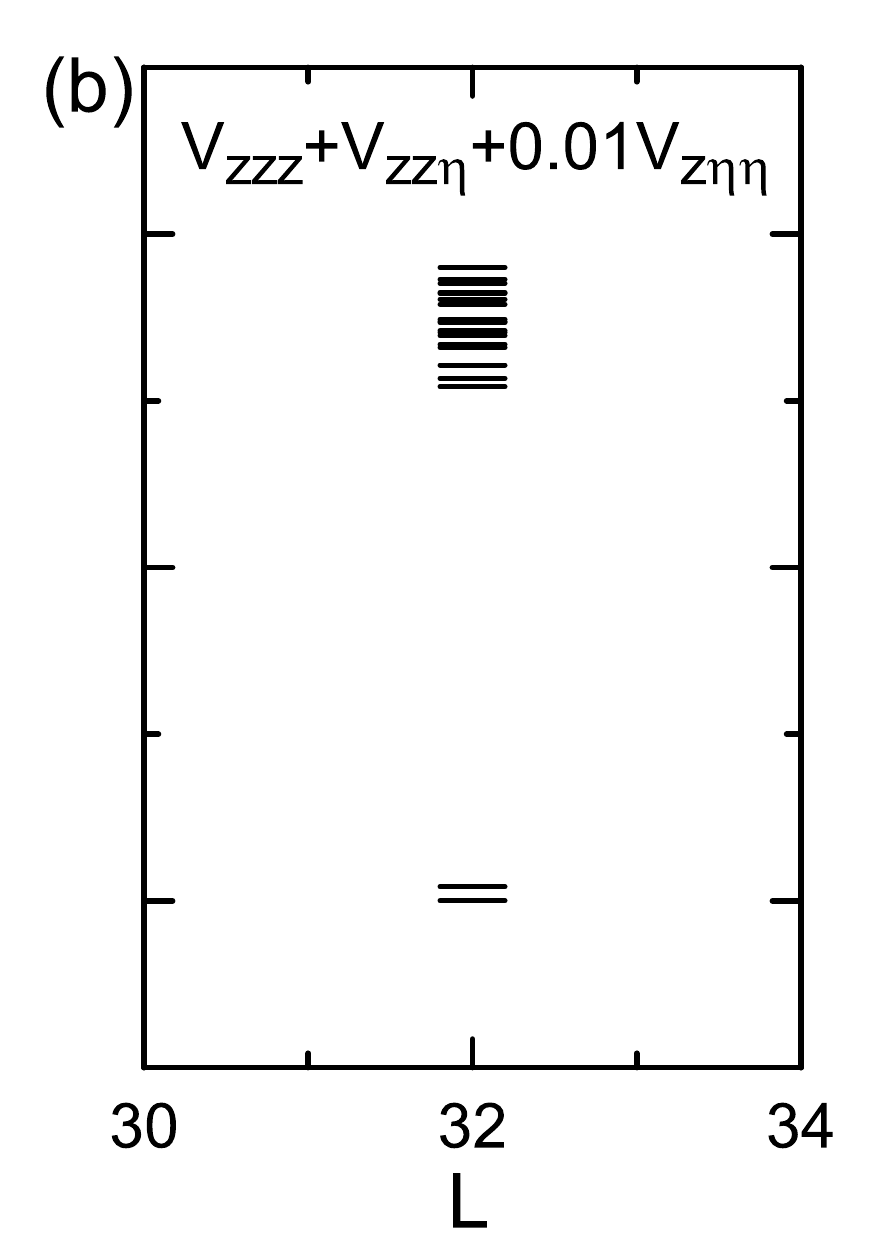}}
\caption{(a) Energy spectrum obtained by diagonalizing the interaction $V=V_{zzz} + V_{zz\eta}$ within a Hilbert space formed by $N=8$ $z$-bosons and two $\eta$-bosons with $L=32$ and $l_{z}^{\text{(max)}} = 12$. There are two zero-energy states in the spectrum. (b) Energy spectrum obtained by adding the $ 0.01 V_{z\eta\eta}$ term to the interaction in (a) and diagonalizing the interaction within the same basis.}
\label{spectra}
\end{figure}

In order to produce $\Psi_{(12)()}$, we observe from the bipartite representation that the interaction 
$V=V_{zzz} + V_{zz\eta}$ annihilates the wave function of $\Psi_{(12)()}$ as well as $\Psi_{(1)(2)}$. We thus diagonalize $V=V_{zzz} + V_{zz\eta}$ in the same basis as above and obtain two zero-energy states in the spectrum, shown in Fig. \ref{spectra} (a). The 2-fold zero-energy subspace is spanned by $\Psi_{(12)()}$ and $\Psi_{(1)(2)}$ because either of the zero-energy states can be expressed as a linear combination of $\Psi_{(12)()}$ and $\Psi_{(1)(2)}$. We then use a small $V_{z\eta\eta}$ interaction to lift the degeneracy. Particularly, we diagonalize the Hamiltonian 
\begin{equation}
V=V_{zzz} + V_{zz\eta} + 0.01 V_{z\eta\eta},
\end{equation}
and obtain a low-energy doublet including one exact zero-energy state and one close-to-zero energy state as shown in Fig. \ref{spectra} (b). The zero-energy state is still exactly $\Psi_{(1)(2)}$, and the close-to-zero energy state approximately corresponds to the orthogonal state in the 2-fold zero-energy subspace. This approximation should become exact when the strength of $V_{z\eta\eta}$ interaction approaches infinitesimally small value. Here for $N=8$, $\Psi_{(12)()}$ is not orthogonal to $\Psi_{(1)(2)}$ though they become orthogonal at thermodynamic limit as discussed in Sec. \ref{sec: backgroundB}. On the other hand, the linear combination $\Psi_{(1)(2)}-2\Psi_{(12)()}$ is essentially orthogonal to $\Psi_{(1)(2)}$ even in relatively small systems as we have shown in Fig. \ref{1} (c). Hence the upper state of the low-energy doublet is essentially the second 2-quasihole qubit state $\Psi_{(1)(2)}-2\Psi_{(12)()}$. In the limit of infinitely small $V_{z\eta\eta}$ and large $N$, the first excited state becomes exactly $\Psi_{(1)(2)}-2\Psi_{(12)()}$. We thus demonstrate that the qubit, namely the two-level system formed by two quasiholes, can be produced with the 3-body Hamiltonian $V=V_{zzz} + V_{zz\eta}$, and each single qubit state can be precisely obtained by including a small $V_{z\eta\eta}$ interaction that lifts the degeneracy between the two states.

\section{Fractional braid statistics \label{sec: statistics}}
\subsection{CFT representation of states}

We briefly outline the derivation of the two qubit states for four quasiholes and their braiding properties. By generalizing the analogy between wave functions for simple FQHE states and chiral correlators in certain conformal field theories, Moore and Read proposed the Pfaffian wave function, which can be written as a correlator within an Ising CFT consisting of three primary fields $1,\psi,\sigma$ of conformal dimensions $0,\frac{1}{2},\frac{1}{16}$ respectively together with a bosonic field $\phi$.~\cite{Moore91}
The real fermion/boson forming an FQHE state is identified as $\psi e^{i\phi \sqrt{m}}$, which is fermionic if $m$ is even and bosonic if $m$ is odd. The Pfaffian wave function made up of $N$ such fermions/bosons can be written as the the correlator of $N$ such operators. The quasihole operator is chosen to be $\sigma e^{i\phi/2\sqrt{m}}$, which satisfies the requirement that the wave function be single-valued in particle coordinates and have the smallest charge ($Q=1/2m$). A wave function with localized quasiholes is obtained by inserting the quasihole operators in the correlator. In particular, the wave function for a 4-quasihole state of a system with (even) $N$ fermion or boson can be written as 
\begin{equation}
\begin{aligned}
\Psi_{\text{4qh}} =& \langle \sigma(\eta_{1}) \cdot\cdot\cdot \sigma(\eta_{4}) \psi (z_{1}) \cdot\cdot\cdot \psi (z_{N}) \\
&\times  e^{ \frac{i } { 2\sqrt{m}} \phi(\eta_{1})} \cdot\cdot\cdot  e^{ \frac{i } { 2\sqrt{m}} \phi(\eta_{4})} \\
&\times e^{i\sqrt{m}\phi(z_{1})} \cdot\cdot\cdot  e^{i\sqrt{m}\phi(z_{N})} e^{-i \int d^{2}z \sqrt{m} \rho_{0} \phi(z) }\rangle.
\label{4qhcorrelation}
\end{aligned}
\end{equation}
The above correlator decomposes into two conformal blocks of $\langle \sigma \sigma \sigma \sigma \psi \cdot\cdot\cdot \psi  \rangle$ corresponding to the two possible ways that the four $\sigma$ operators can fuse to give $1$ under the fusion rules of the Ising CFT.\cite{DiFrancesco97} The fields $\sigma(\eta_1) \sigma(\eta_2)$ and the fields $\sigma(\eta_3) \sigma(\eta_4)$ can each fuse to a $1$ or $\psi$ fields and the two choices result in the two wave functions $\Psi^1$ and $\Psi^\psi$. Exact form of the conformal blocks can be obtained through bosonization.\cite{Nayak96,ardonne2010chiral}. The explicit wave function given in Eq. (\ref{Nayak}) can be obtained with the help of the identity~\cite{DiFrancesco97}:
\begin{equation}
\langle e^{i\alpha_{1} \phi(z_{1})} \cdot\cdot\cdot e^{i\alpha_{N} \phi(z_{N})} \rangle = \prod_{i<j} (z_{i} - z_{j})^{\alpha_{i} \alpha_{j}}.
\end{equation}
Charge neutrality required in the above equation is ensured by the smeared background charge.~\cite{Moore91} It has been conjectured~\cite{Nayak96} that if the wave functions of states are written in the basis specified by the conformal blocks, the Berry phase vanishes - all braiding properties are manifested in the basis functions. This can be understood using the plasma analogy~\cite{Gurarie97,Bonderson11} and has been numerically verified using Monte-Carlo methods~\cite{Tserkovnyak03}. Thus we can now read off the braiding properties directly from the basis wave functions. As we have discussed in Sec. \ref{sec: backgroundB}, the 4-quasihole qubit states $\Psi^{(1,\psi)}$ reduce to the two 2-quasihole states [shown in Eq. (\ref{2qh1andpsi})] if two of the quasiholes are sent to infinity. Since the Berry phase matrix vanishes in this basis, the phase gained when braiding $\eta_{1}$ around $\eta_{2}$ should be equal to 
\begin{eqnarray}
{2\pi\over 8} \;\; &{\rm for}&\;\; \Psi_{(1)(2)} \label{phase1} \\
{10\pi\over 8} \;\; &{\rm for}&\;\; \Psi_{(1)(2)}-2\Psi_{(12)()} \label{phase2}
\end{eqnarray}
Here, $\Psi_{(1)(2)}$  and $\Psi_{(12)()}$ describe the quasihole states rather than the states with two species of bosons. However, since bosons do not produce any additional Berry phases (modulo $2\pi$) for exchanges or windings, the bound state of a boson and a quasihole should obey the same fractional statistics as the quasihole. We verify these results numerically in the following subsections.

\subsection{Effective relative angular momenta \label{sec: angularmomenta}}

The possibility of fractional relative angular momentum was introduced in a previous work\cite{Zhang14} in the context of quantum Hall states at filling fractions $\nu=\frac{n}{n\pm1}$, which we briefly review here. The relative angular momentum of a pair of bosonic particles in the lowest Landau level (LLL) is quantized to even numbers. It was shown that if the same bosons bind anyonic quasiholes of a quantum Hall droplet, the boson-anyon bound states exhibit \emph{fractional} effective relative angular momenta whose values are directly related to the Abelian fractional statistics of the anyons. It was demonstrated that the binding of a boson to an anyonic excitation can be achieved with a contact interaction between the auxiliary bosons and the majority bosons forming the quantum Hall droplet. It was shown that the fractional angular momentum results in a ``quantization" of the separation $R$ between the auxiliary bosons given by the expression $R^2\sim 2M_{\rm eff}{\ell}^{'2}$ where $M_{\rm eff}=M-\nu$ and $\ell^{'2}=\frac{\ell^2}{1-\nu}$, which may be observable in measurements of the pair correlation function for the auxiliary bosons.

The combined wave function of the two auxiliary bosons and the background quantum Hall droplet takes the form 
\begin{eqnarray}
\psi(z,\eta_{1},\eta_{2})\sim&&\Psi_\nu(z) \times \prod_{i=1}^N (z_i-\eta_2)(z_i-\eta_1)\times P_M(\eta_1,\eta_2) \nonumber\\
&&\times e^{-\sum_{i=1} ^N \frac{|z_i|^2}{4\ell^2} - \frac{|\eta_1|^2+|\eta_2|^2}{4\ell^2} }
\label{boson-anyon-wave function}
\end{eqnarray}
where $\Psi_\nu(z)$ is the wave function of the homogeneous incompressible quantum Hall state at filling $\nu$
and  
\be
P_{M} (\eta_{1}, \eta_{2})=\sum_{L={\rm even}} C_L (\eta_1-\eta_2)^L (\eta_1+\eta_2)^{M-L}
\ee
is a symmetric polynomial with total angular momentum $M$. In our evaluations of the pair correlation function we will assume the most symmetric situation in which $\eta_2$ lies at the origin. Then, with $\eta\equiv\eta_{1}$ denoting the relative coordinate, the wave function reduces to 
\begin{eqnarray}
\psi(z,\eta) \sim\Psi_\nu(z) \prod_{i=1}^N [z_i(z_i-\eta)] \eta^M 
 e^{-\sum_{i=1} ^N \frac{|z_i|^2}{4\ell^2} - \frac{|\eta|^2}{4\ell^2} }
\label{boson-anyon-wave2}
\end{eqnarray}
where $M$ can take all integer values.

The relative distance between the bosons in such a wave-function can be estimated using a simple semi-classical picture in which the bosons can be thought of as revolving around each other at a distance $R$. The most probable value of $R$ is fixed by the requirement that the total Aharanov-Bohm phase accumulated along an orbit of radius $R$ be equal to the Berry phase gained by the wave-function upon winding a boson around the other at the same radius. The latter corresponds to the number of vortices that $\eta_1$ sees on the other bosons enclosed by the orbit. Using a mean field assumption of uniform distribution of bosons in the quantum Hall droplet away from the quasiholes, we can arrive at the above mentioned dependence of the radius $R$. 

The above argument can be rephrased in terms of a conjecture that the quantum Hall quasihole wave functions when written as the conformal blocks of appropriate CFTs have a normalization constant that is independent of the locations of the quasiholes~\cite{Gurarie97}. For example, consider the 2-quasihole wave function at filling fraction $\nu=1/q$ of bosons:
\begin{eqnarray}
\psi_{\rm CFT,\frac{1}{q}} (z,\eta_1,\eta_2)\propto (\eta_1-\eta_2)^\frac{1}{q} \prod_{i=1}^N\prod_{j=1,2} (z_i-\eta_j) \nonumber\\ 
\times \prod_{i<j=1}^N (z_i-z_j)^q e^{-\sum_{i=1}^N \frac{|z_i|^2}{4\ell^2}-\frac{ |\eta_1|^2+|\eta_2|^2}{4q\ell^2}}.
\end{eqnarray}
For sufficiently separated quasihole locations, the conjecture is that the normalization $\mathcal{N}=\int dz |\psi_{\rm CFT,\frac{1}{q}}(z,\eta_1,\eta_2)|^2$ is weakly dependent on $\eta_{1,2}$. In terms of this, the pair distribution function $G(\eta_1,\eta_2)$ of the combined wave function of the two species of bosons in Eq. (\ref{boson-anyon-wave function}) is given by
\begin{eqnarray}
G(\eta_1,\eta_2) \propto&& \int dz |\psi(z,\eta_{1},\eta_{2})|^2\nonumber \\
&& = |P_{M}(\eta_{1},\eta_{2})|^{2} |\eta_1-\eta_2|^{-\frac{2}{q}} \nonumber \\
&& \quad \times \mathcal{N} e^{-(1-\frac{1}{q})\frac{ |\eta_1|^2+|\eta_2|^2}{2\ell^2}}.
\end{eqnarray}
The peak of the pair correlation function $G(\eta_1=0,|\eta_2|=R)$ is then located at
\begin{equation}
R^2 = 2(M-\nu)\frac{\ell^2}{1-\nu}.
\end{equation}

This derivation of the relation between $M$ and $R^2$ can be extended to the case of the 2-quasihole wave functions of the Pfaffian. In the above example of the Abelian quantum Hall state, the 2-quasihole state created in the quantum Hall droplet due to the repulsion between the two species of bosons corresponded to a unique wave function with two single-quantum vortex excitations.  The Pfaffian presents the possibility of a qualitatively different behavior. For a suitably chosen interaction between the species (as discussed in the previous section), the Pfaffian quantum Hall wave function produces half-quantum vortex excitations at the locations of the auxiliary bosons. Unlike the Abelian case, here there are two degenerate states with such quasiholes.

For the MR quasiholes, the wave function in Eq. (\ref{boson-anyon-wave function}) is replaced by 
\begin{eqnarray}
\psi(z,\eta_1,\eta_2) \sim \Psi_{\rm 2qh} (z,\eta_1 , \eta_2) \times P_{M}(\eta_1,\eta_2)
\label{boson-nonabelion-wave function}
\end{eqnarray}
where $\Psi_{\rm 2qh} (z,\eta_1,\eta_2)$ could be an arbitrary state in the two-dimensional space of 2-quasihole states 
\begin{equation}
\Psi_{\rm 2qh} (z,\eta_1,\eta_2) = a\Psi_{(1)(2)} +  b(\Psi_{(1)(2)}-2\Psi_{(12)()}).
\end{equation}
The only dependence of $\Psi_{\rm 2qh} (z,\eta_{1},\eta_{2})$ on $\eta_{i}$'s is through the wave functions $\Psi_{(1)(2),(12)()}$ and not through coefficients $a$ and $b$. In terms of normalizations $\mathcal{N}^{1,\psi}$ of $\Psi^{(1,\psi)}_{\rm 2qh}$ (shown in Eq. (\ref{2qh1andpsiappendix})), which by conjecture are weakly dependent on $\eta_{1,2}$, we can write the pair correlation $G(\eta_1,\eta_2)$ in the wave function in Eq. (\ref{boson-nonabelion-wave function}) as:
\begin{eqnarray}
G(\eta_1,\eta_2) \propto [ |a|^2 |P_{M}(\eta_{1},\eta_{2})|^{2} |\eta_1-\eta_2|^{-\frac{2}{8}} \mathcal{N}^1 + \nonumber \\
 |b|^2 |P_{M}(\eta_{1},\eta_{2})|^{2} |\eta_1-\eta_2|^{-\frac{10}{8}} \mathcal{N}^\psi ] \times e^{(\frac{1}{2}-1)\frac{|\eta_1|^2+|\eta_2|^2}{2\ell^2}}.
\label{normalization}
\end{eqnarray}
The cross term between $\Psi^{(1)}_{\rm 2qh}$ and $\Psi^{(\psi)}_{\rm 2qh}$ is proportional to the overlap between the two states and has been assumed to be zero in arriving at the above result. Maximization of the pair correlation function $G(0,R)$ with respect to $R^2$ as before gives 
\begin{equation}
4\left(M-\frac{5}{8}\right) \leq R^2\leq 4\left(M-\frac{1}{8}\right) 
\label{nonAbelianMeff}
\end{equation}
where the right and left hand limits correspond to the $(a,b)=(0,1)$ and $(1,0)$ respectively. 
The fractional part of the effective relative angular momentum, as can be read out in Eq. (\ref{nonAbelianMeff}), exactly gives the Berry phases $(1/8)2\pi$ and $(5 /8)2\pi$ in Eqs. \ref{phase1} and \ref{phase2}.

The relation in Eq. (\ref{nonAbelianMeff}) can also be derived within a simple semiclassical approximation mentioned above for quantum Hall states with Abelian anyons. For 2-quasihole Pfaffian states, one needs to consider an additional statistical phase $\delta^{(1,\psi)}$ that $\eta_{1}$ sees on $\eta_{2}$ associated with different fusion channels. We assume $\eta_{2}$ is fixed at the origin, and  equate the number of flux quanta ($\phi_0=hc/e$) enclosed by the circle of radius $R$, i.e. $\pi R^2 B/\phi_0=R^2/2\ell^2$, to the mean number of enclosed vortices $\frac{1}{2}\tilde{N}+M+\delta^{(1,\psi)}$. Here $\tilde{N} =  \nu R^{2}/2\ell^{2} -\frac{1}{2}\nu$ is the average number of enclosed $z$ particles, and the $\frac{1}{2}$ factor in front of $\tilde{N}$ comes from the fact that the quasihole of Pfaffian state has only half-quantum vortex; the last term $-\frac{1}{2}\nu$ reflects the expulsion effect of the centrally located quasihole. Solving for $R^{2}$, we obtain
\begin{equation}
R^2 = 4(M+\delta^{(1,\psi)} -\nu/4)  \ell^{2}
\label{nonAbelianMeff2}
\end{equation}
with $\nu=1$ for bosonic Pfaffian state. The phases $\delta^{(1,\psi)}$ cannot be estimated from such a semi-classical picture but can be seen to be $\delta^{(1)}=1/8$ and $\delta^{(\psi)}=-3/8$ by comparison with the limiting cases of Eq. (\ref{nonAbelianMeff}). 

The mean field derivation of Eq. (\ref{nonAbelianMeff2}) above has assumed the condition that the $\eta$-particle is inside the disk of quantum Hall state of $z$. For larger M, $\eta_{1}$ is pushed outside the quantum Hall droplet, and the number of enclosed vortices is simply $M+N/2+\delta^{(1,\psi)}$, which leads to the relation
\begin{equation}
R^2=2(M+N/2+\delta^{(1,\psi)}) \ell^2\;\;{\rm (outside)}.
\label{outside}
\end{equation}

\subsection{Numerical studies}

\begin{figure}
\resizebox{0.45\textwidth}{!}{\includegraphics{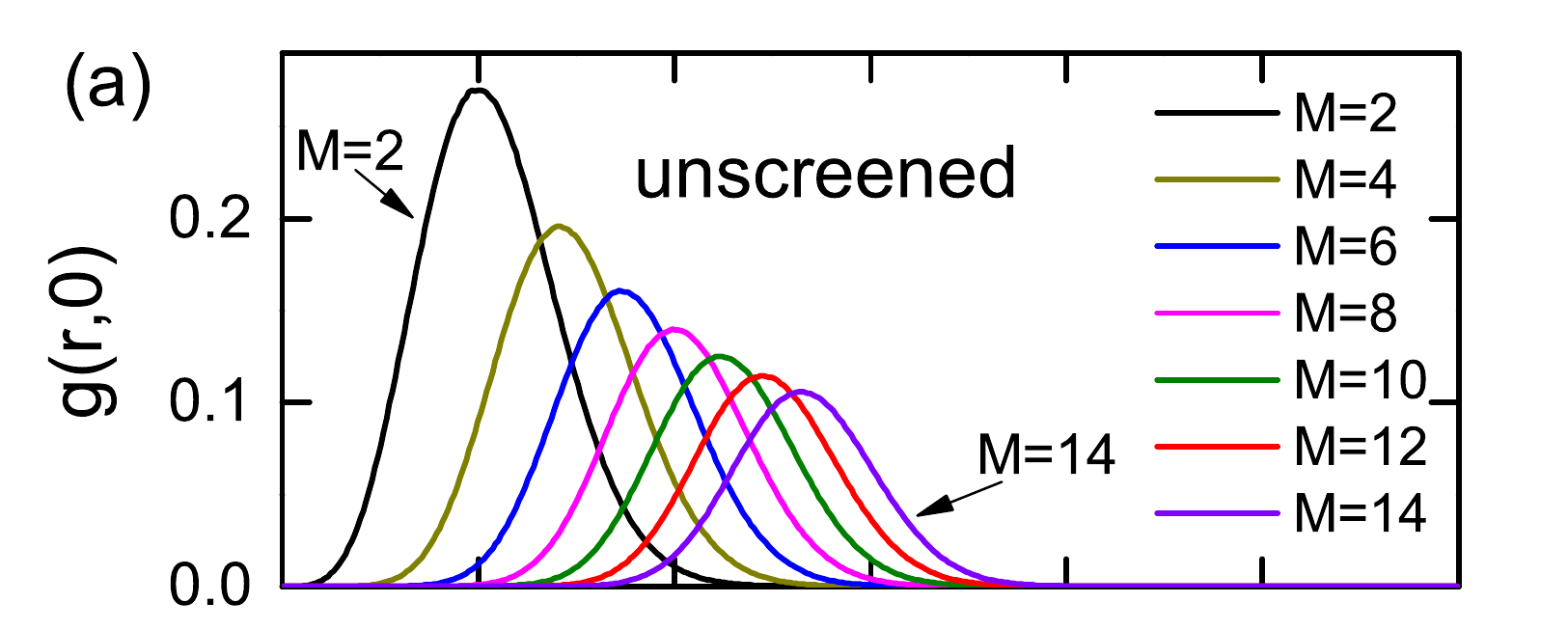}}\\
\vspace{-3mm}
\resizebox{0.45\textwidth}{!}{\includegraphics{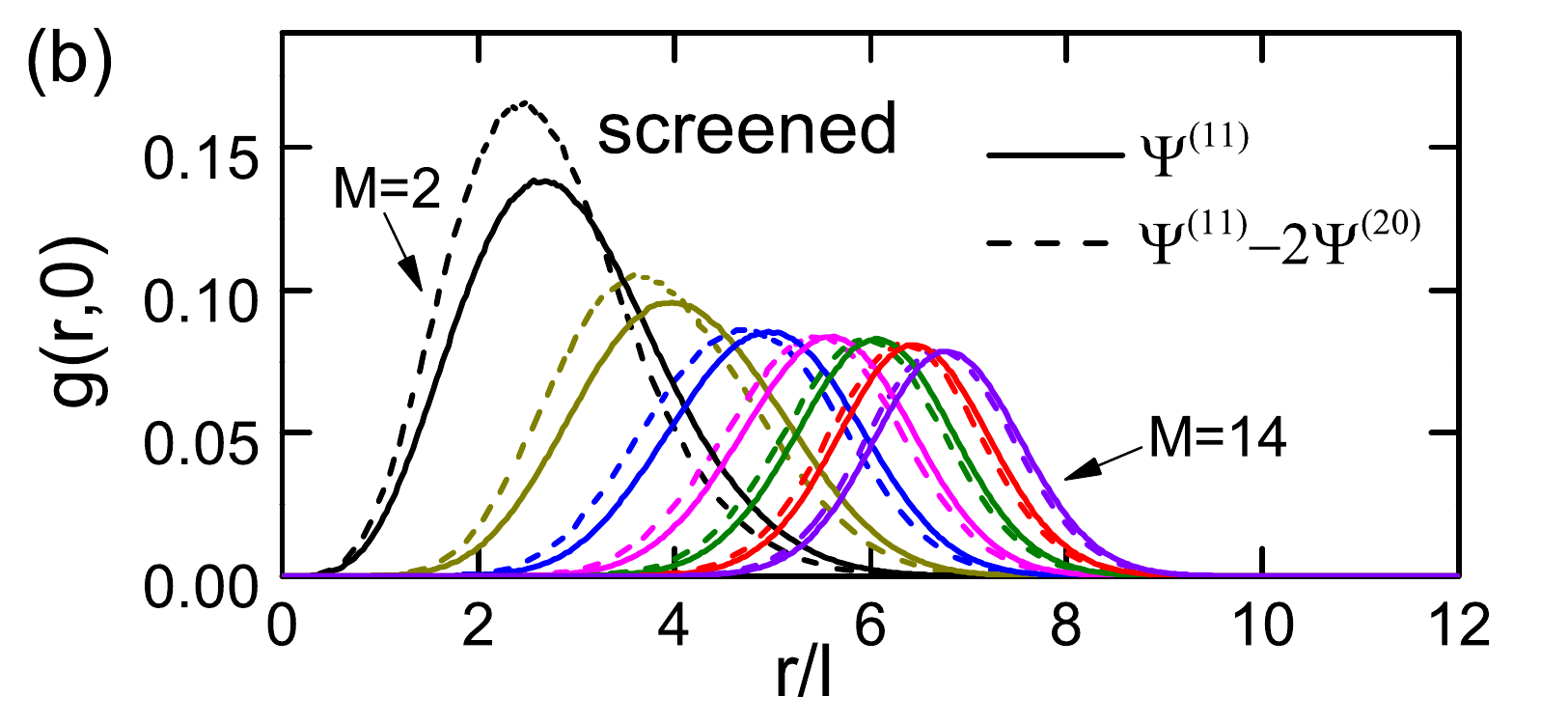}}
\caption{(Color online). The pair correlation function $g(r,0)$ of the auxiliary bosons with one boson fixed at the center (a) without any background $z$ particles and (b) in the background of a $\nu=1$ bosonic Pfaffian state of $N=20$ $z$ particles. The solid and dashed lines in (b) show the pair correlation function for the different 2-quasihole states $\Psi_{(1)(2)} P_{M}(\eta_{1},\eta_{2})$ and $(\Psi_{(1)(2)}-2\Psi_{(12)()}) P_{M}(\eta_{1},\eta_{2})$ respectively. The pair correlation functions here and in the following are all quoted in units of $\rho_{0}=(2\pi\ell^{2})^{-1}$, where $\ell$ is the magnetic length.}
\label{density}
\end{figure}

We next verify the two limiting cases in Eq. (\ref{nonAbelianMeff}) with explicit calculation of the pair correlation functions $g(|\eta_{1}| = r, \eta_{2} = 0)$ for states specified by $(a,b)=(0,1)$ and $(1,0)$ respectively. We fix one auxiliary boson at the origin, and numerically calculate the density profile of the other one with Monte-Carlo techniques. As a reference, we first consider a system containing only the two auxiliary bosons $\eta_{1}$ and $\eta_2$ for $2<M<14$ (with no $z$ bosons), in which case the peak of the pair correlation function is simply given by $R^2 =2M \ell^2$.~\cite{Zhang14} The result is shown in Fig. \ref{density} (a). As we introduce the two auxiliary bosons into a bosonic Pfaffian state of $z$-bosons, the density profile of the bosons screened by the quantum Hall fluid ``moves outwards'' as shown in Fig. \ref{density} (b), where the solid line displays the pair correlation for $\Psi_{(1)(2)}$ state and the dashed line for $(\Psi_{(1)(2)}-2\Psi_{(12)()})$ state. The change in the length scale governing the auxiliary pair of bosons ($\ell^2\to 2\ell^2$) is mainly due to the reduced effective magnetic field caused by the interaction between auxiliary bosons and background bosons.~\cite{Zhang14} 

\begin{figure}
\hspace{-5mm}
\resizebox{0.25\textwidth}{!}{\includegraphics{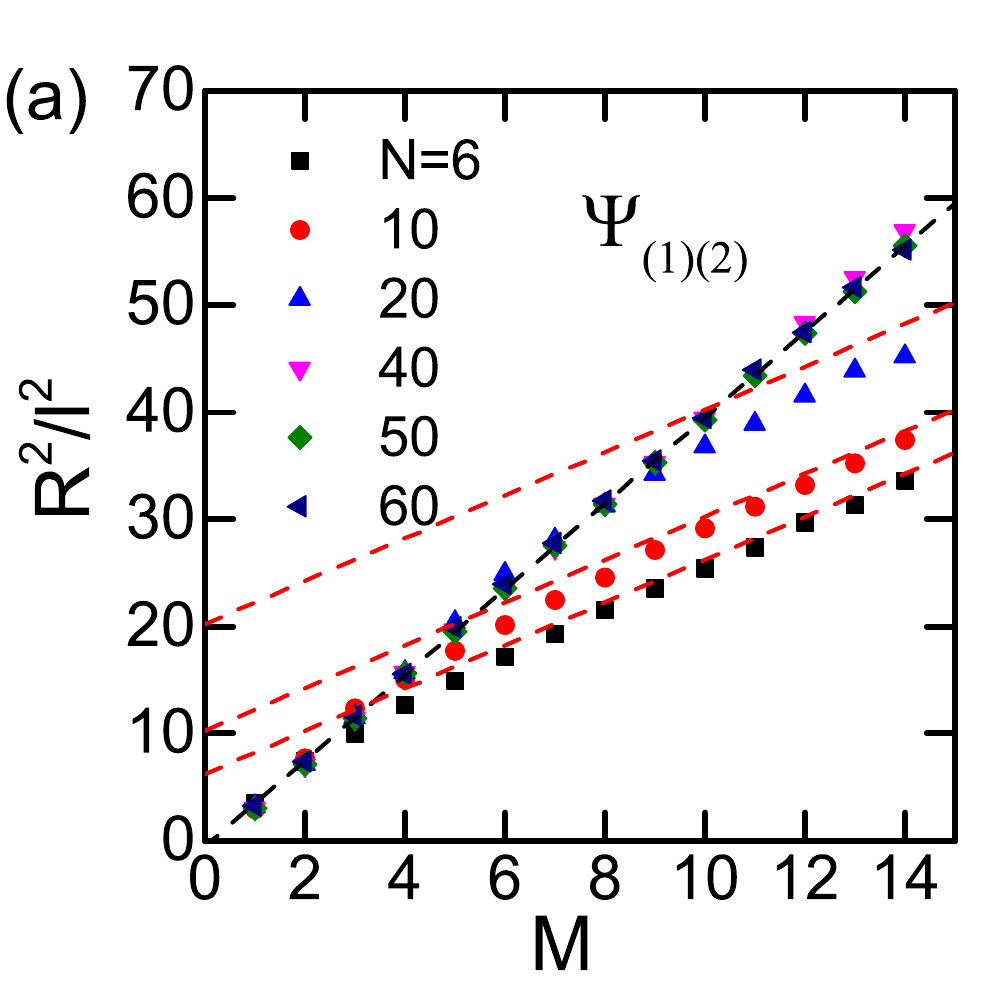}}
\hspace{-3mm}
\resizebox{0.25\textwidth}{!}{\includegraphics{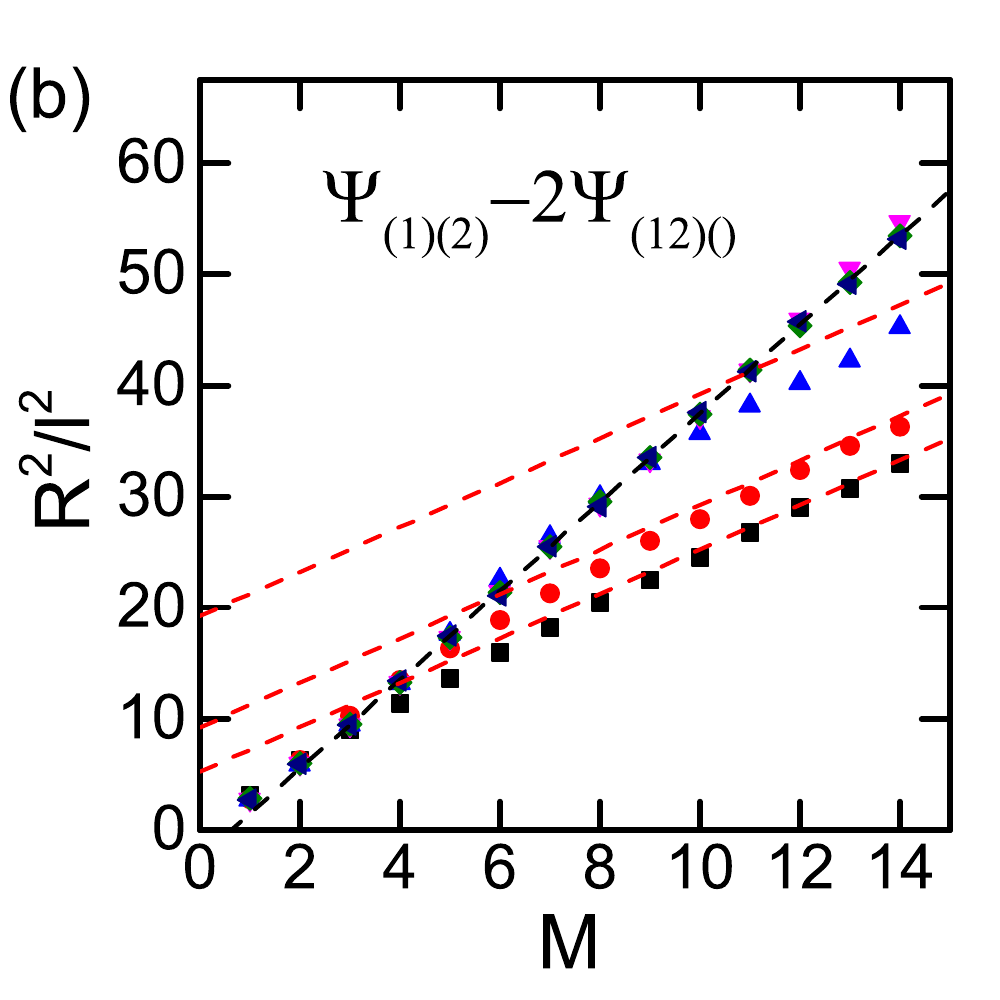}}\\
\hspace{-4.5mm}
\resizebox{0.255\textwidth}{!}{\includegraphics{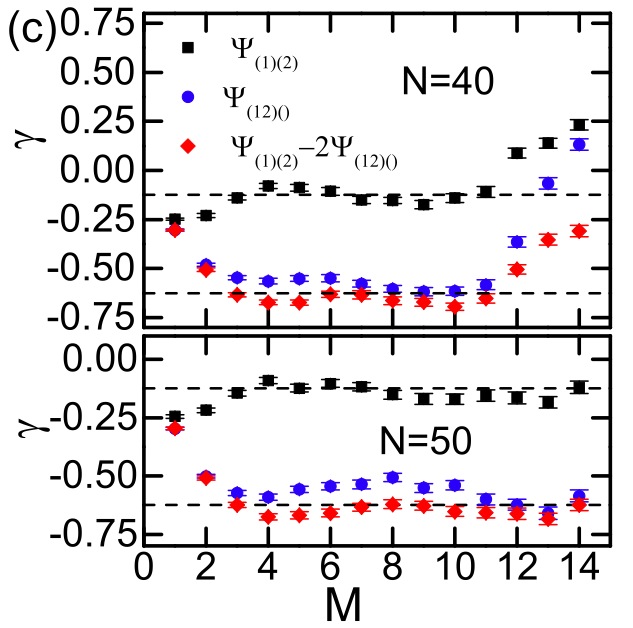}}
\hspace{-3.4mm}
\resizebox{0.255\textwidth}{!}{\includegraphics{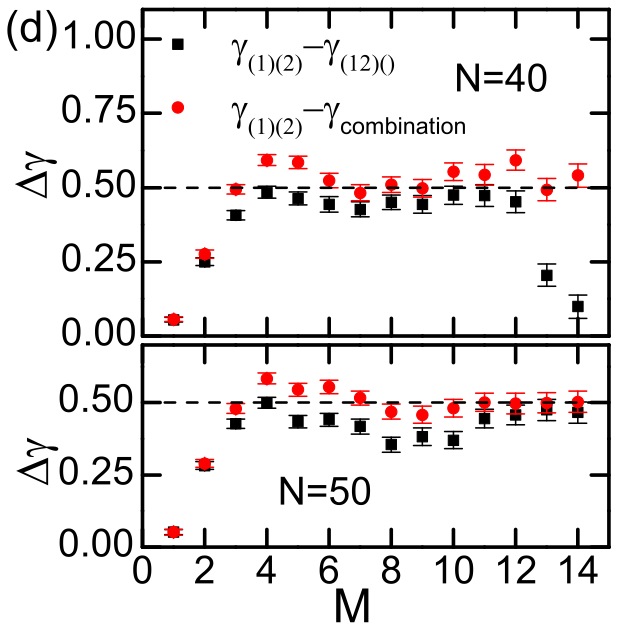}}
\caption{(Color online). The peak positions of the pair correlation function as a function of the relative angular momentum $M$ for $N=6-60$, for (a) $\Psi_{(1)(2)}$ state and (b) $\Psi_{(1)(2)}-2\Psi_{(12)()}$ state. The black dashed line plots the predicted behavior in Eq. (\ref{nonAbelianMeff2}) and the red dashed line plots Eq. (\ref{outside}). (c) The quantity $ \frac{1}{4}(R^{2}/l^{2} - 4M)$ as a function of M for $N=40$ and $N=50$ for $\Psi_{(1)(2)}$ (black square), $\Psi_{(12)()}$ (blue circle), and $\Psi_{(1)(2)}-2\Psi_{(12)()}$ (red diamond). The black dashed lines plot the expected $\gamma$ value: $-1/8$ and $5/8$. (d) The differences of the quantities $ \frac{1}{4}(R^{2}/l^{2} - 4M)$ between $\Psi_{(1)(2)}$ and $\Psi_{(12)()}$ are shown in the black square, and between $\Psi_{(1)(2)}$ and $\Psi_{(1)(2)}-2\Psi_{(12)()}$ are shown with the red circle.}
\label{Rpeak}
\end{figure}

To further explore the fractional statistics from the pair correlation function, we plot the square of the peak positions as a function of $M$ and $N$ in Figs. \ref{Rpeak}(a) and \ref{Rpeak}(b) for the two states $\Psi_{(1)(2)}$ and $(\Psi_{(1)(2)}-2\Psi_{(12)()})$. The predicted behaviors from Eqs. (\ref{nonAbelianMeff2}) and (\ref{outside}) are also shown with black and red dashed lines for comparison. The behavior predicted by Eq. (\ref{nonAbelianMeff2}) is fully confirmed for quasiholes in the interior of the quantum Hall droplet. The behavior changes from Eq. (\ref{nonAbelianMeff2}) to Eq. (\ref{outside}) for large $M$, when the auxiliary boson $\eta_{1}$ lies outside the droplet. Note that the fractional part of the effective angular momentum $\gamma=\delta^{(1,\psi)} - \nu/4$ is just given by the x-intercept of the black dashed line. To study $\gamma$ more precisely, we write Eq. (\ref{nonAbelianMeff2}) as $R^{2}/\ell^{2} = 4(M+\gamma)$ and solve for $\gamma$: $\gamma = \frac{1}{4}(R^{2}/l^{2} - 4M)$. We therefore plot the quantity $ \frac{1}{4}(R^{2}/l^{2} - 4M)$ as a function of M for $N=40$ and $50$ in Fig. \ref{Rpeak} (c). This quantity is displayed for the three wave functions:  $\Psi_{(1)(2)}$, $\Psi_{(12)()}$ and $\Psi_{(1)(2)}-2\Psi_{(12)()}$, and the expected values, $\gamma = -1/8$ for  $\Psi_{(1)(2)}$ and $\gamma  = -5/8$ for  $\Psi_{(1)(2)}-2 \Psi_{(12)()}$, are also plotted with dashed line. Figure \ref{Rpeak}(c) shows that the numerical result of $\gamma$ is consistent with the expected value, and the state $\Psi_{(12)()}$ (corresponding to $(a,b)=(1,-1)$) produces a $R^{2}$ value in between the two limits in Eq. (\ref{nonAbelianMeff}). The deviation at small $M$ ($M=1,2$) is unsurprising because quasiholes need to be well separated to exhibit the expected braid statistics \cite{Kjonsberg99,Kjonsberg99b,Jeon03b}, and the deviation at larger $M$ for $N=40$ indicates proximity of $\eta_{1}$ to the edge. Figure \ref{Rpeak}(d) plots the differences of the quantities in (c) between  $\Psi_{(1)(2)}$ and $\Psi_{(12)()}$ (black square), and between $\Psi_{(1)(2)}$ and $\Psi_{(1)(2)}-2\Psi_{(12)()}$ (red circle). The latter one is expected to be $0.5$. 

From these numerical results, we stress that even though it is not easy to detect the accurate fractional Berry phase from small system ($N=6 \cdots 10$),  the pair correlation of the auxiliary pair has an obviously different profile for the two topologically different states especially for small $M$. Both of these profiles are also clearly distinguishable from that of  an unscreened pair. The radius of a small screened pair with $M=2$ is close to the expected value $2 \sqrt{2+\gamma}\ell$ even for small system size $N=6$, which is significantly different from the radius $R=2\ell$ of the unscreened pair. We note that the smallest pair with $M=1$ is not meaningful for the purposes of braid statistics, presumably due to the non-negligible overlap between the two quasiholes. Therefore, in our discussion below of the experimental feasibility in Sec. \ref{sec: exp}, we will only consider $M \geq 2$.

\section{Many quasiholes \label{sec: manyquasihole}}

We have shown that the 2-quasihole wave functions can be produced in a system with two species of bosons interacting with a 3-body potential. Wave functions for states containing $2n$ quasiholes, in principle, can also be produced as low/zero energy states of a system of $N$ $z$-bosons and $2n$ auxiliary bosons under a 3-body interaction Hamiltonian. Specifically, one can see from the bipartite form of the MR $2n$-quasihole states (shown in Eq. (\ref{bimanyqh})) that they have zero energy for the interaction $V=V_{zzz} + V_{zz\eta}$. As we describe in the following, the zero-energy states of this Hamilonian have degeneracies that exceed the $2^{n-1}$-fold degeneracy of the FQHE quasihole states, due to the additional degree of freedom associated with the wave function of the auxiliary bosons. However when these degrees of freedom are frozen - for example by fixing the locations of the auxiliary bosons to produce localized quasiholes, the expected $2^{n-1}$ degeneracy is obtained.

The uninteresting degeneracies arising from bosonic edge excitations of the quantum Hall system can be removed by constraining the edge with a momentum cut off. We diagonalize the interaction $V=V_{zzz} + V_{zz\eta}$ for a system of $N=6$ $z$-bosons and four auxiliary bosons, with a cut-off on single-particle angular momentum equal to $l_{\text{max}} = N$ (corresponding to the standard MR 4-quasihole wave function $\Psi_{\text{4qh}}$ in a form of Eq. (\ref{4qh-1})). For simplicity we study only those sectors that have the same total angular momentum as the MR quasihole state. Numerically we find that the energy spectrum has three zero-energy states. 

The counting can be explained by noting that the zero-energy states should have the form $\Psi(z,\eta) \times g(\eta)$ where $\Psi(z,\eta)$ is a $2n$-quasihole state of the FQH system and is annihilated by $V_{zzz}$. Taking $\{\eta_{i}\}$ as dynamic coordinates for the auxiliary bosons, $\Psi(z,\eta)$ is also annihilated by $V_{zz\eta}$ because of the binding of the quasihole to the bosons. The function $g(\eta)$ is any possible wave function of the $2n$ auxiliary bosons.  A basis for space of such zero-energy states can be obtained as  $\varphi_{F,\lambda}(z,\eta) \times h_{\mu}(\eta)$ where $\varphi_{F,\lambda}$ are the quasihole basis states introduced in Eq. (\ref{RRformOfQHs}). $h_{\mu}(\eta)$ are wave functions of $2n$ bosons which can be indexed by a partition (non-decreasing ordered sequence) $\mu$ of length $2n$ and has a form $h_{\mu}(x) = \mathcal{S}[x_1^{\mu_1} x_2^{\mu_2}\dots]$. The total angular momentum of this state is 
\begin{equation}
L_{F,\lambda,\mu} = \frac{N^2}{2} + N(n-1) + \sum_{k=1}^F \lambda_k - \frac{F}{2}(2n-1)+ \sum_{l=1}^{2n} \mu_{l}
\end{equation}
Those states that have the same total angular momentum $\frac{N^2}{2} + N(n-1)$ as the MR state of $2n$ quasiholes satisfy the constraint $\sum_{k=1}^F \lambda_k+ \sum_{l=1}^{2n} \mu_{l} = \frac{F}{2}(2n-1)$. Correct degeneracy can be estimated by enumerating the states that satisfy this constraint. For the above mentioned case of $2n=4$ quasiholes, the states are
\begin{align}
(F=0,\lambda=[\;]{}&,\mu=[0,0,0,0])\nonumber\\
(F=2,\lambda=[0,1]{}&,\mu=[0,0,0,2])\nonumber\\
(F=2,\lambda=[0,1]{}&,\mu=[0,0,1,1])\nonumber
\end{align}
In order to produce localized quasiholes, we fix the values of $\eta_{i}$'s, and then the degeneracy of the zero-energy states reduces to the expected number $2^{n-1}=2$. 
It can be explicitly checked that the space spanned by the $2^{n-1}$ as obtained states is identical to the space of the $2^{n-1}$ MR quasihole states with same quasihole positions.

The same calculation for six quasiholes is performed with $L=N^{2}/2 +2N$ and  $l_{\text{max}}=N+1$. There are $11$ zero-energy states for $6$ $\eta_{i}$-bosons immersed in $N=6$ $z$-bosons, and the degeneracy reduces to $4$ when fixing the $\eta_{i}$ positions. The degeneracy of $11$ can again be obtained by enumerating $(F,\lambda,\mu)$ that satisfy the mentioned constraint.
 
We summarize the degeneracy results in Table \ref{table1} and conclude that the MR $2n$-quasiholes states can be produced by turning on a 3-body interaction  $V=V_{zzz} + V_{zz\eta}$ in a system of $2n$ auxiliary bosons plus $N$ $z$-bosons and destroying the auxiliary bosons at some particular positions, leaving $2n$ quasiholes at those locations. 

For larger values of $l_{\rm max}$, the number of zero energy states is greater due to edge excitations. We have not studied the counting of such states.

\begin{table}
\caption{Number of zero-energy states for $2-6$ quasiholes}
\begin{ruledtabular}
\begin{tabular}{c|c|c|c}
number of & $l_{\rm max}$ & degeneracy for &  degeneracy for \\ 
quasiholes $2n$ &  &   bosons& localized quasiholes
\\ \hline
2 & $N-1$  & 1 & 1 \\
4 & $N$ &3 & 2\\
6 & $N+1$ & 11 & 4
\end{tabular}
\end{ruledtabular}
\label{table1}
\end{table}

\section{Feasibility in ultra-cold atom systems \label{sec: exp}}

\subsection{Preparing the model wave functions}

FQHE states can in principle exist in rapidly rotating gases of ultra-cold bosonic atoms. By an adiabatic scheme implemented by Gemelke {\em et al.}\cite{Gemelke10}, one can reach the ground state at a given total angular momentum $L$ for $N$ bosons. We have discussed in the Sec. \ref{sec: Hamiltonian} that the 2-quasihole qubit states can be produced with a 3-body Hamiltonian. We now study how a Pfaffian wave function of the form $\Psi_{\text{2qh}} P_{M}(\eta_{1}, \eta_{2})$ can be prepared. We will only consider $M\geq 2$ cases as we discussed above that $M=1$ is too small to show meaningful information. The standard Hamitonian in the LLL with a 3-body interaction (in the rotating frame) is
\begin{equation}
H=V_0\sum_{i,j<k}^{N+2} \delta(\vec{r}_{i}-\vec{r}_{j})\delta (\vec{r}_{i}-\vec{r}_{k})+(\omega_z-\Omega) \hat{L}_z+(\omega_{\eta}-\Omega)  \hat{L}_{\eta}, 
\label{H}
\end{equation}
where $\Omega$ is the rotation frequency, $\omega_{z}$ and $\omega_{\eta}$ are the harmonic confinement frequencies, and $L_{z}$ and $L_{\eta}$ are total angular momentum operators for the $z$ and $\eta$ particles respectively. We assume the interaction is independent of the species which produces the $\Psi_{(1)(2)}$ state as its unique zero-energy state at $L_{0}=N^{2}/2$.  As shown above, the other 2-quasihole state $\Psi_{(1)(2)}-2\Psi_{(12)()}$ can be produced as the first excited state for an appropriately chosen interaction. For $L=L_{0} + M$,  $\Psi_{(1)(2)} P_{M}(\eta_{1}, \eta_{2})$ has zero interaction energy, but it is in general not the only zero-energy state; other zero-energy states can be constructed wherein the additional angular momentum $M$ is absorbed by $z$ particles. We ask if the edge excitations of the $z$ particles can be suppressed by taking $\omega_{z} \gtrsim \omega_{\eta}$, following the strategy used in 
Ref. \onlinecite{Zhang14}. The exact ground state of this Hamiltonian at $M=2$ and $M=3$ is well approximated by $\Psi_{(1)(2)} P_{M}(\eta_{1}, \eta_{2})$, while for $M\geq4$ the second or third excited state has a pair correlation function matching well with that of $\Psi_{(1)(2)} P_{M}(\eta_{1}, \eta_{2})$. At minimum, the state of $M=2$ and $M=3$ can be adiabatically prepared in principle. Then the measurement for pair correlation function can be implemented through single-atom detection combined with a short time-of-flight expansion. More detailed discussion can be found in Ref. \onlinecite{Zhang14}.

\subsection{2-body contact interaction}

We have shown above that many relevant topological states with two or more quasiholes can be engineered with 3-body interaction. However, the 3-body interaction is not easy to implement.  A more realistic interaction in cold atom systems is the 2-body contact interaction:
\begin{equation}
H=V_0\sum_{i<j} \delta(\vec{r}_{i}-\vec{r}_{j}).
\end{equation}
The MR Pfaffian state is not the exact ground state of this Hamiltonian. Encouragingly, however, it has been shown~\cite{Wilkin00, Cooper01, Regnault03, Regnault07, Chang05b} that this interaction produces an incompressible state at $\nu=1$, which is accurately, though not exactly, described by the Pfaffian wave function. We therefore ask to what extent the above results carry over to the 2-body contact interaction. We focus on the following two aspects: (i) Can a 2-body interaction produce the low-energy Pfaffian quasihole states with the expected degeneracy? (ii) In order to detect statistics through pair correlation function measurement, we ask whether a 2-body interaction can produce a ground state of which the pair correlation function for two auxiliary bosons is similar to that of the Pfaffian wave function $\Psi_{\text{2qh}} P_{M}(\eta_{1}, \eta_{2})$. We will only discuss the $\Psi_{(1)(2)}$ state for simplicity because similar conclusions apply to  $\Psi_{(1)(2)} -2\Psi_{(12)()}$. We will also focus on $M\geq2$ cases as $M=1$ has been shown ineffective in reflecting statistics information correctly.

To answer the first question, we perform exact diagonalization of 6-10 bosons with an additional 2 and 4 auxiliary bosons within a Hilbert space specified by a total angular momentum $L$ and a cut-off on single-particle angular momentum $l_{\rm max}$. The value of $L$ is fixed according to the Pfaffian quasihole wave functions, for example, $L=N^{2}/2$ for 2-quasihole states $\Psi_{(1)(2)}$ and $\Psi_{(12)()}$. The value of $l_{\rm max}$ is taken to be equal or greater than the largest single-particle angular momentum in wave functions.
In order to tune the 2-body interaction, we write the interaction potential as 
\begin{equation}
V=V_{zz}+\lambda_{1}V_{z\eta}+\lambda_{2}V_{\eta\eta}, 
\end{equation}
where $\lambda_{1}$ and $\lambda_{2}$ control the relative strength of $z$-$\eta$ and $\eta$-$\eta$ interactions. Because the 3-body interaction $V=V_{zzz} + V_{zz\eta}$ excluding $\eta$-$\eta$ interaction term ($V_{z\eta\eta}$) produces degenerate Pfaffian quasihole states, we keep $\lambda_{2}=0.0$ and only tune the value of $\lambda_{1}$ for the 2-body interaction. We have not been able to find any values of $\lambda_{1}$ and $l_{\rm max}$ that produce a degenerate or quasi-degenerate set of low-energy states with significant overlap with the Pfaffian quasihole wave functions, for either two or four quasiholes.

\begin{figure}
\resizebox{0.43\textwidth}{!}{\includegraphics{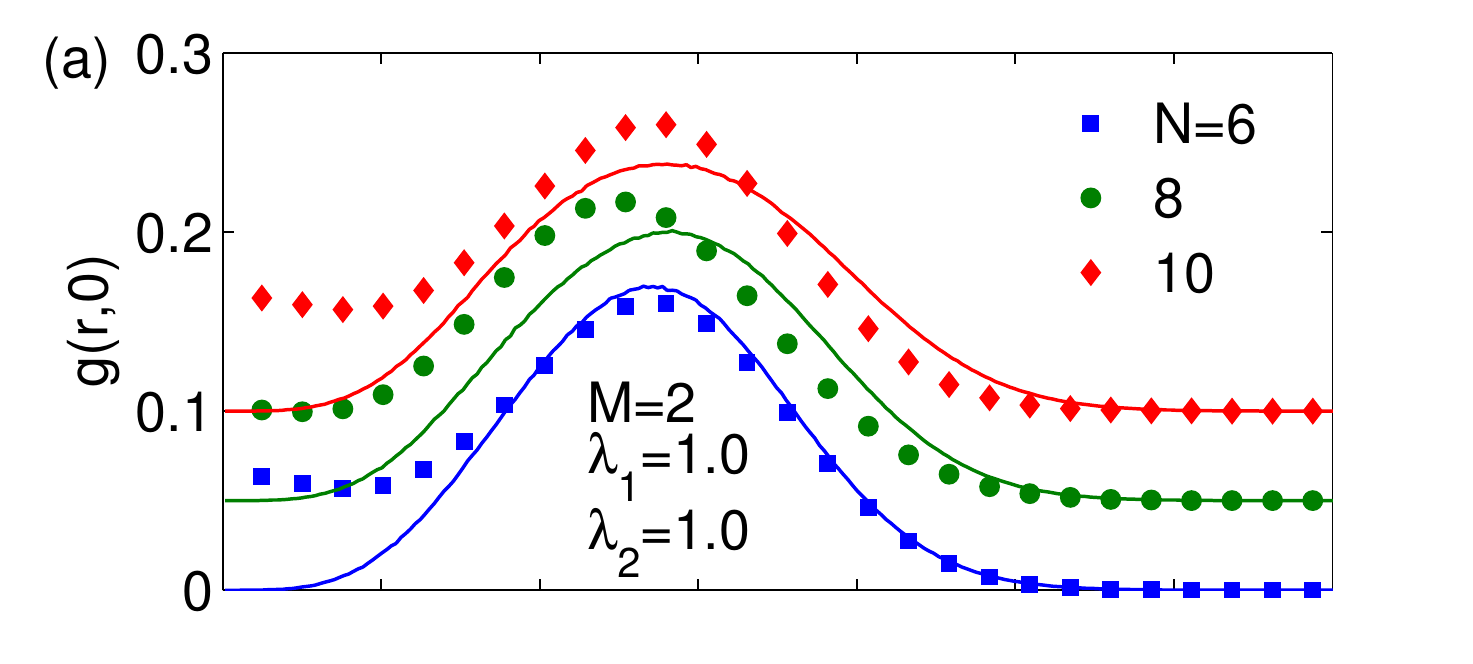}}\\
\vspace{-3.0mm} 
\resizebox{0.43\textwidth}{!}{\includegraphics{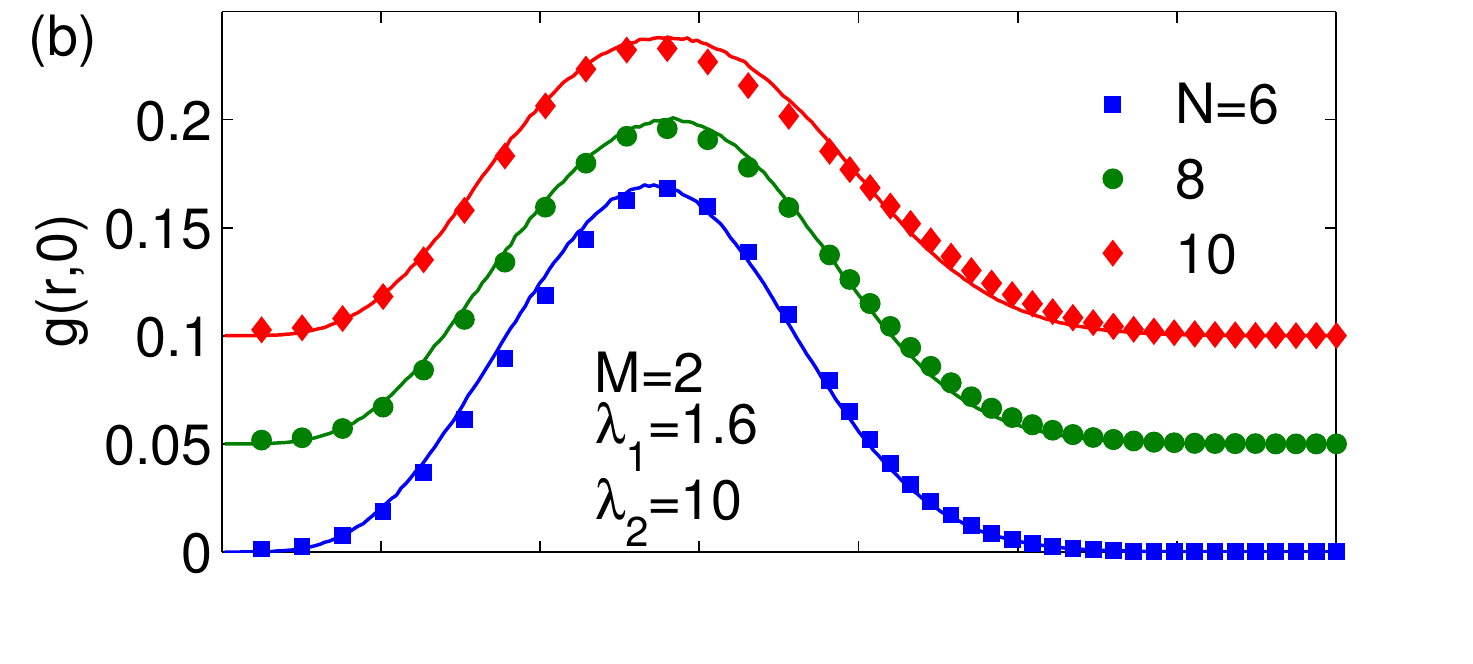}}\\
\vspace{-4.8mm} 
\resizebox{0.434\textwidth}{!}{\includegraphics{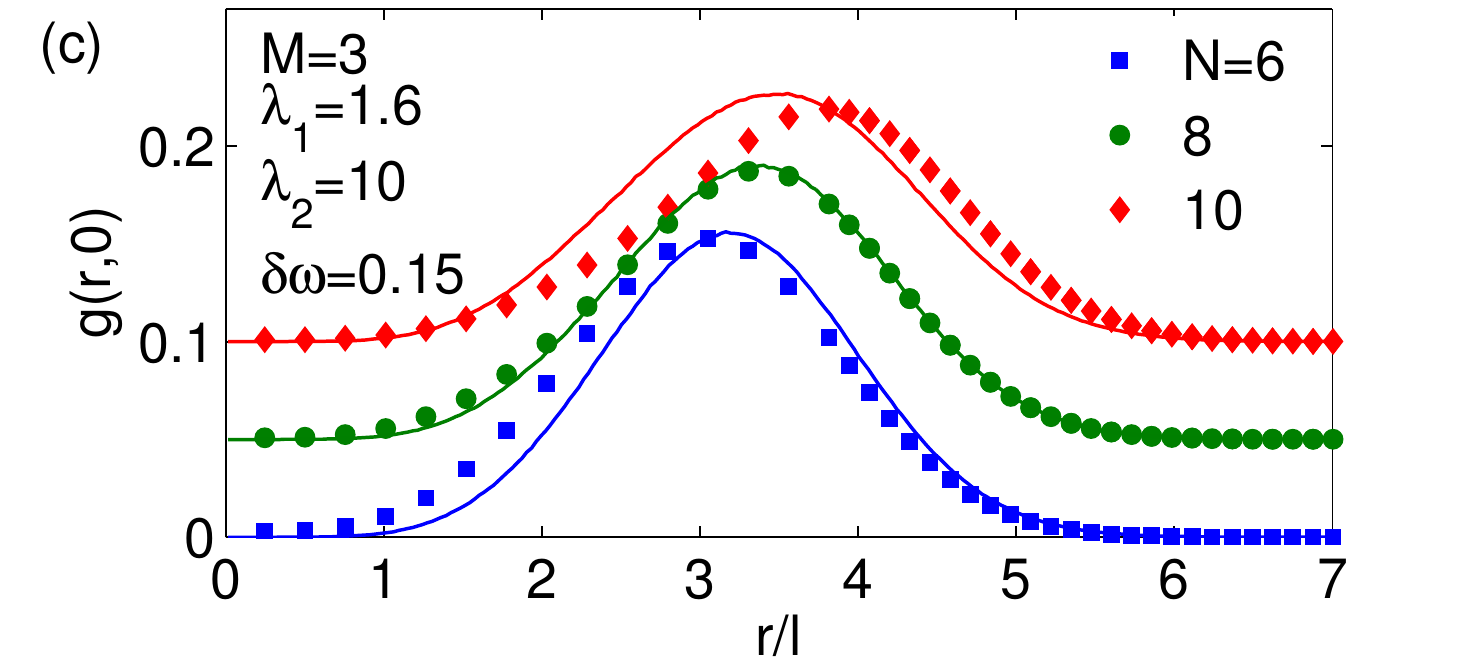}}
\caption{(Color online). The pair correlation functions $g(r,0)$ of the $\eta$-bosons of the ground state obtained from diagonalizing the 2-body Hamiltonian with parameters shown in the figure are plotted for different system sizes $N=6$ (blue square), $N=8$ (green circle), $N=10$ (red diamond). As comparison, the pair correlation functions calculated from the model wave functions $\Psi_{(1)(2)} P_{M}(\eta_{1}, \eta_{2})$ are also shown with solid lines. The plots for $N=8$ and $10$ have been shifted up by $0.05$ and $0.1$ units respectively for ease of depiction. (a)(b) For $M=2$, an isotropic contact interaction fails to produce a similar pair correlation function as that of the model wave function, but it is possible to find a universal set of interaction parameters $\lambda_{1,2}$ for different system sizes that gives agreement. (c) For $M=3$, a slightly different confinement potential for $z$- and $\eta$-bosons is necessary to produce a desired ground state. However, the proper value of parameter $\delta\omega$ sensitively depends on the system size. }
\label{2body}
\end{figure}

Now we study the second question for $M=2$. We recall that the Pfaffian wave function $\Psi_{(1)(2)} P_{M}(\eta_{1}, \eta_{2})$ can be approximately produced as the ground state of the Hamiltonian in Eq. (\ref{H}) with $\omega_{z} \gtrsim \omega_{\eta}$  at $L=N^{2}/2+M$. In fact, if we take $l_{\rm max}$ to be the minimal value $l_{\rm max}=N-1$, the zero-interaction-energy states have the exact unique form $\Psi_{(1)(2)} P_{M}(\eta_{1}, \eta_{2})$, without requiring the confinement term in the Hamiltonian. We now diagonalize an isotropic 2-body interaction ($\lambda_{1}=\lambda_{2}=1.0$) with the same $l_{\rm max}=N-1$ and $L=N^{2}/2+M$ for $M=2$. The pair correlation functions for the $\eta$-bosons of the ground state are clearly different from that of the wave function $\Psi_{(1)(2)} P_{M}(\eta_{1}, \eta_{2})$, as shown in Figure \ref{2body} (a). We next tune the parameters $\lambda_{1,2}$ and find that $\lambda_{2}=10$ and $\lambda_{1}=1.6$ give good agreement for system size of $N=8$, which also apply to $N=6$ and $N=10$ as shown in Figure \ref{2body} (b).

For $M>2$, a confinement potential that is slightly stronger for $z$-bosons than  $\eta$-bosons needs to be included in the Hamiltonian in order to help the $\eta_{i}$'s absorb all of the additional angular momentum to produce the multiplicative factor $ P_{M}(\eta_{1}, \eta_{2})$. We denote the difference in the harmonic confinement frequencies of the two species of bosons as $\delta\omega=\omega_{z} -\omega_{\eta}$. By diagonalizing the new Hamiltonian with different $\delta\omega$ at $L=N^{2}/2+M$ and $l_{\rm max}=N-1$, we find that the pair correlation function of the ground state depends sensitively on the value of $\delta\omega$. Figure \ref{2body} (c) shows the pair correlation function of the ground state obtained with parameters $\lambda_{1} =1.6$, $\lambda_{2} =10$ and $\delta\omega=0.15$ at $M=3$ for different $N$. It is not possible to find a single set of parameters that gives the desired pair correlation function for all $N$.

Up to now we have studied the $\Psi_{(1)(2)}$ state; the same arguments apply to the other state  $\Psi_{(1)(2)}- 2\Psi_{(12)()}$ with optimal parameters $\lambda_{1}=0.4$, $\lambda_{2}=10$ and $\delta\omega=0.023$ for $N=8$. In the study of $M=2,3$ cases, all exact diagonalizations are performed with the smallest value of $l_{\rm max}$ as specified by the wave function. A larger $l_{\rm max}$ generally requires a different set of parameters. In conclusion, 2-body interaction fails to produce a realization of the Pfaffian quasihole wave functions with the desired degeneracy, at least for all the systems we have considered. 

\section{Conclusions \label{sec: conclusion}}

We have pursued the idea that the non-Abelian quasiholes in the Pfaffian fractional quantum Hall states can be captured and manipulated by introducing auxiliary bosons of a different species. We have shown how, assuming a precise control of the Hamiltonian, this can be used to produce qubit states with desired topological properties. Furthermore, a measurement of the pair distribution function of the auxiliary bosons can provide a direct confirmation of the Abelian and non-Abelian braid statistics of these quasiholes. 

{\bf Acknowledgments}: This work was supported by the U.S. Department of Energy, Office of Science, Basic Energy Sciences, under Award No. DE-SC0005042. We thank Eddy Ardonne, Jerome Dubail, Nate Gemelke, and Ying-Hai Wu for valuable discussions. We thank Research Computing and Cyberinfrastructure at Pennsylvania State University (supported in part through instrumentation funded by the National Science Foundation through grant OCI-0821527) and MPI-PKS for providing computing resources.

\appendix
\section{Equivalence of Bipartite and Pfaffian Representations of $2n$-quasihole states \label{ap1}}
We demonstrate that the Pfaffian wave function of $2n$-quasihole states can be represented as an equivalent bipartite form. We first consider the 2-quasihole states for simplicity. There are two 2-quasihole states in the bipartite representation: one in which the two quasiholes are constructed in the same partition and another one in which the two quasiholes are constructed in different partitions. They have the form:
\begin{widetext}
\begin{equation}
\Psi_{(12)()}(z,\eta_1,\eta_2)= \mathcal{S}\left\{ \prod_{i=1}^{\frac{N}{2}}(z_{2i}-\eta_1)(z_{2i}-\eta_2) \prod_{i<j=1}^{\frac{N}{2}} (z_{2i}-z_{2j})^2 (z_{2i-1}-z_{2j-1})^2  \right\},
\end{equation}

\begin{equation}
\Psi_{(1)(2)}(z,\eta_1,\eta_2)=\mathcal{S}\left\{ \prod_{i=1}^{\frac{N}{2}}(z_{2i}-\eta_1)(z_{2i-1}-\eta_2) \prod_{i<j=1}^{\frac{N}{2}} (z_{2i}-z_{2j})^2 (z_{2i-1}-z_{2j-1})^2  \right\}.
\label{A2}
\end{equation}
\end{widetext}
Following, we demonstrate that these states can be represented in terms of the Pfaffian of a matrix. We will take $\Psi_{(12)()}$ as an example and the proof for $\Psi_{(1)(2)}$ can be completed by simply replacing $z_{2i}$ in the factor $(z_{2i} - \eta_{2})$ with $z_{2i-1}$ throughout the derivation.

Factorizing out a  Jastrow factor $J(z)=\prod_{i<j}^{N} (z_{i} -z_{j})$ from the wave function gives
\begin{widetext}
\begin{equation}
\begin{aligned}
\Psi_{(12)()}(z,\eta_1,\eta_2)&=J(z)\times\mathcal{A}\left\{ \prod_{i=1}^{\frac{N}{2}}(z_{2i}-\eta_1)(z_{2i}-\eta_2) \frac{\prod_{i<j=1}^{\frac{N}{2}} (z_{2i}-z_{2j}) (z_{2i-1}-z_{2j-1}) }{\prod_{i,j=1}^{\frac{N}{2}} (z_{2i}-z_{2j-1}) } \right\} \\
&=(-1)^{\frac{N}{4}(\frac{N}{2}-1)}  J(z) \times \mathcal{A}\left\{ \prod_{i=1}^{\frac{N}{2}}(z_{2i}-\eta_1)(z_{2i}-\eta_2)  \text{det}\frac{1}{z_{2i}-z_{2j-1}} \right\} \\
&=(-1)^{\frac{N}{4}(\frac{N}{2}-1)} J(z) \times \sum_{\sigma \in S_{N/2}} \mathrm{sgn}(\sigma) \mathcal{A}\left\{ \prod_{i=1}^{\frac{N}{2}}(z_{2i}-\eta_1)(z_{2i}-\eta_2)  \prod_{i=1}^{\frac{N}{2}} \frac{1}{z_{2i}-z_{2\sigma_{i}-1}} \right\}.
\end{aligned}
\label{proof1}
\end{equation}
\end{widetext}
The symbol $\mathcal{A}$ denotes antisymmetrization over all coordinates $z_{i}$, and an identity due to Cauchy~\cite{MacDonald79}, 
\be
\text{det}\frac{1}{z_{2i}-z_{2j-1}} = (-1)^{\frac{N}{4}(\frac{N}{2}-1)}  \frac{\prod_{i<j}^{\frac{N}{2}} (z_{2i} - z_{2j}) (z_{2i-1}-z_{2j-1})} {\prod_{i,j}^{\frac{N}{2}} (z_{2i} - z_{2j-1}) }
\ee
has been used to obtain the second line of Eq. \ref{proof1}. The third line comes from the expansion of the determinant where the sum is computed over all permutations $\sigma$ of the set $\{1, 2, ..., N/2\}$. Note that all the permutations yield the same contribution, thus the wave function can be simplified as
\begin{widetext}
\begin{equation}
\begin{aligned}
\Psi_{(12)()}(z,\eta_1,\eta_2) &=(-1)^{\frac{N}{4}(\frac{N}{2}-1)}(N/2)!  J(z) \times\mathcal{A} \prod_{i=1}^\frac{N}{2} \frac{(z_{2i}-\eta_1)(z_{2i}-\eta_2)}{z_{2i}-z_{2i-1}}  \\
&=(-1)^{\frac{N}{4}(\frac{N}{2}-1)}(N/2)!  J(z) \times \mathrm{Pf} \left( \frac{ (z_{i}-\eta_{1})(z_{i}-\eta_{2})+ (i \leftrightarrow j)}{z_{i}-z_{j}} \right),
\end{aligned}
\label{proof2}
\end{equation}
\end{widetext}
which is identical to the Pfaffian representation in Eq. (\ref{2qh20pf}) up to an overall normalization factor. The definition of Pfaffian has been used in the last step, and the Gaussian factors are suppressed for simplicity.

The above derivation can be easily generalized to $2n$-quasihole state by replacing the factor $\prod_{i=1}^{\frac{N}{2}}(z_{2i}-\eta_1)(z_{2i-1}-\eta_2)$ in Eq. (\ref{A2}) with $\prod_{i=1}^{\frac{N}{2}} [(z_{2i}-\eta_\alpha)(z_{2i}-\eta_\beta)... (z_{2i-1}-\eta_\rho)(z_{2i-1}-\eta_\sigma)...]$. Thus a bipartite representation for the Pfaffian wave function in Eq. (\ref{manyqh}) is constructed:
\begin{widetext}
\begin{equation}
\Psi_{(\alpha\beta...)(\rho\sigma...)}=\mathcal{S}\left\{\prod_{i=1}^{\frac{N}{2}}(z_{2i}-\eta_\alpha)(z_{2i}-\eta_\beta)... (z_{2i-1}-\eta_\rho)(z_{2i-1}-\eta_\sigma)... \prod_{i<j=1}^{\frac{N}{2}} (z_{2i}-z_{2j})^2 (z_{2i-1}-z_{2j-1})^2  \right\}.
\label{bimanyqh}
\end{equation}
\end{widetext}

\section{2-quasihole representation of the 4-quasihole qubit states in certain limit \label{ap2}}
We now show that the 2-quasihole states $\Psi_{(1)(2)} $ and $\Psi_{(1)(2)}-2\Psi_{(12)()} $ can represent  the two 4-quasihole qubit states in Eq. (\ref{Nayak}) in the limit that two of the quasiholes are sent to infinity. We take $\eta_{3}=r, \eta_{4} = r e^{i\theta}$ and $r \rightarrow \infty$ for Eq. (\ref{Nayak}). Now $\Psi_{(13)(24)} $ and $\Psi_{(14)(23)} $ reduce to the standard MR 2-quasihole wave function $\Psi_{(1)(2)}$ with quasiholes located at $\eta_{1}$ and $\eta_{2}$, while $\Psi_{(12)(34)} $ reduce to $\Psi_{(12)()}$, with a same normalization factor for all of them as shown below:
\begin{equation}
\begin{aligned}
\Psi_{(13)(24)}, \Psi_{(14)(23)} \rightarrow r^{N} e^{\frac{N\theta}{2}} \Psi_{(1)(2)},\\
\Psi_{(12)(34)} \rightarrow r^{N} e^{\frac{N\theta}{2}} \Psi_{(12)()}.
\end{aligned}
\end{equation}

In this limit, the factor $x \rightarrow 1+ (1-e^{-i\theta}) (\frac{\eta_{1} - \eta_{2}}{r})  = 1 +\epsilon (\eta_{1} - \eta_{2})$, where $\epsilon = \frac{1-e^{-i\theta}}{r}$ is a very small number. Before we apply any approximation, we rewrite $\Psi_{(13)(24)} \pm \sqrt{x}\Psi_{(14)(23)}$ in Eq. (\ref{Nayak}) as 
\begin{equation}
 (x \pm \sqrt{x})\Psi_{(14)(23)} + (1-x)\Psi_{(12)(34)} 
\end{equation}
with the help of the identity 
\begin{equation}
\Psi_{(12)(34)} -\Psi_{(14)(23)} = \frac{1}{x} (\Psi_{(12)(34)} -\Psi_{(13)(24)})
\end{equation}
from Ref. \onlinecite{Nayak96}. By applying $\sqrt{x} \rightarrow 1 + \frac{1}{2} \epsilon (\eta_{1} - \eta_{2})$, the two topologically different 4-quasiholes wave functions $\Psi^{(1,\psi)} $ in  Eq. (\ref{Nayak}) are written as (for $m=1$): 
\begin{widetext}
\begin{equation}
\Psi^{(1,\psi)} \rightarrow \frac{(\eta_{1} - \eta_{2})^{\frac{1}{8}}  } {(1\pm (1+\frac{1}{2} \epsilon (\eta_{1} - \eta_{2} ) )^{1/2}} ( (x \pm \sqrt{x})\Psi_{(14)(23)} + (1-x)\Psi_{(12)(34)} ).
\end{equation}
Thus 
\begin{equation}
\Psi^{(1)} \rightarrow (\eta_{1} - \eta_{2})^{1/8} \Psi_{(14)(23)} \rightarrow (\eta_{1} - \eta_{2})^{1/8} \Psi_{(1)(2)}
\label{B5}
\end{equation}
and 
\begin{equation}
\begin{aligned}
\Psi^{(\psi)} &\rightarrow \frac{(\eta_{1} - \eta_{2})^{\frac{1}{8}}  } {(-\frac{1}{2} \epsilon (\eta_{1} - \eta_{2} ) )^{1/2}}[\frac{1}{2}\epsilon(\eta_{1} - \eta_{2})(\Psi_{(14)(23)} - 2\Psi_{(12)(34)}) ] \rightarrow (\eta_{1} - \eta_{2})^{5/8} ( \Psi_{(1)(2)}-2\Psi_{(12)()} )
\end{aligned}
\label{B6}
\end{equation}
\end{widetext}
Therefore $\Psi^{(1)}$ reduces to $\Psi_{(1)(2)}$, and $\Psi^{(\psi)} $ reduces to a linear combination of the two kinds of 2-quasihole wave functions $(\Psi_{(1)(2)}-2\Psi_{(12)()})$, with prefactors $(\eta_{1} - \eta_{2})^{1/8} $ and $(\eta_{1} - \eta_{2})^{5/8} $ respectively.

In the above derivation we have neglected the factor $e^{-  \sum_{\alpha=1}^{4} \frac{|\eta_{\alpha}|^{2}}{8m\ell^{2}}}$ in Eq. (\ref{Nayak}). This factor simply contributes to an overall normalization constant of the wave function and is only considered when the dependence of the normalization constant on the locations of quasiholes is discussed, as in the derivation of Eq. (\ref{normalization}) in Sec. \ref{sec: angularmomenta}. In that case, a more precise version of Eq. (\ref{B5}) and (\ref{B6}) is needed as
 \begin{equation} 
\begin{aligned}
\Psi^{(1)} & \rightarrow & \Psi^{(1)}_{\rm 2qh} &= (\eta_{1} - \eta_{2})^{1/8} \Psi_{(1)(2)} e^{- \frac{|\eta_{1}|^{2} + |\eta_{2}|^{2}}{8\ell^{2}}}
\\
\Psi^{(\psi)} & \rightarrow & \Psi^{(\psi)}_{\rm 2qh}  &= (\eta_{1} - \eta_{2})^{5/8} ( \Psi_{(1)(2)}-2\Psi_{(12)()} ) e^{- \frac{|\eta_{1}|^{2} + |\eta_{2}|^{2}}{8\ell^{2}}}.
\end{aligned}
\label{2qh1andpsiappendix}
\end{equation}

\bibliography{biblio_fqheold.bib}
\bibliographystyle{apsrev4-1}

\end{document}